\documentclass[aoas]{imsart}

\RequirePackage{amsthm,amsmath,amsfonts,amssymb}
\RequirePackage[authoryear]{natbib}
\RequirePackage[colorlinks,citecolor=blue,urlcolor=blue]{hyperref}
\RequirePackage{graphicx}
\RequirePackage{booktabs}
\usepackage{xcolor}

\startlocaldefs

\theoremstyle{plain}

\theoremstyle{definition}

\newcommand{\manualref}[1]{#1}

\endlocaldefs

\makeatletter
\def\journal@name{}
\makeatother

\begin{document}

\begin{frontmatter}
\title{Decomposition Atlases of Brains and Kidneys from Postmortem CT: Shape and Topological Data Analysis}
\runtitle{Decomposition Atlases of Postmortem Brains and Kidneys}

\begin{aug}

\author[A]{\fnms{Yanyan}~\snm{Zhan}}
\author[A]{\fnms{Yuexuan}~\snm{Wu}}
\author[B]{\fnms{Ian L.}~\snm{Dryden}}
\author[C]{\fnms{Zeloso M.}~\snm{Tiago}}
\author[C]{\fnms{Edward C.}~\snm{Schwalbe}}


\address[A]{Department of Statistics, University of South Carolina, USA}

\address[B]{School of Mathematical Sciences, University of Nottingham, UK}

\address[C]{School of Geography and Natural Sciences, Northumbria University, UK}



\end{aug}

\begin{abstract}


Forensic assessments of postmortem decomposition often rely on external, whole-body signs, yet structural changes within organs remain largely unquantified. Using whole-body computed tomography scans of 59 decedents from the New Mexico Decedent Image Database, we conduct an organ-level statistical shape analysis of decomposing brains and kidneys, representing each organ as a three-dimensional point cloud of its outer surface and internal cavity boundaries. Decomposition creates voids, fragmentation, and structural heterogeneity, complicating point correspondence. Registration-based principal component analysis and principal nested cones yield interpretable orderings for brains only. We therefore construct a registration-free pipeline using persistent homology and a fusion of complementary topological distances selected through simulations of cavity formation and growth. Clustering identifies broad patterns of internal decay, and seriation produces organ-specific decomposition atlases. The brain atlas combines geometric and topological rankings; the kidney atlas uses topology alone. Rank agreement is higher between paired kidneys (Kendall’s $\tau = 0.59$) than between brains and kidneys ($\tau = 0.34$–$0.39$), indicating asynchronous decompositions within a body. Whole-body stages agree more closely with brain than with kidney rankings. These findings support quantitative forensic taphonomy by complementing whole-body staging with organ-level, interior-based measurements. We release the organ point clouds and analysis code to support research on organ decomposition and shape analysis without reliable correspondence.
\end{abstract}


\end{frontmatter}

\section{Introduction}

Assessing how far a body has decomposed is a routine but consequential task in forensic practice: the assessment informs estimates of the postmortem interval, interpretation of autopsy findings, and decisions about identification. However, postmortem decomposition is a complex process influenced by intrinsic and extrinsic factors and their interactions \citep{zhou2011factors}. Routine staging commonly relies on externally visible changes to the head, trunk, and limbs, as in the widely used total body score of \citet{megyesi2005using} and its refinements \citep{moffatt2016improved}, or draws on ecological indicators \citep{zhou2011factors}. However, these scores summarize a complex, organ-dependent process into a limited number of externally observable traits, and they rest on an implicit assumption that external appearance is an adequate proxy for internal state.

Postmortem computed tomography (PMCT) permits non-invasive examination of internal anatomy and is an established complement to autopsy \citep{roberts2012post}. Imaging-based decomposition measures, however, provide limited morphological detail. For example, the radiological alteration index grades putrefactive gas at a small number of anatomical sites but does not describe the morphology of the organs themselves \citep{egger2012development}. Quantitative postmortem imaging research has largely focused on hard tissue, which persists longer and is easier to segment than decomposed soft tissue. Internal-organ morphology has received less quantitative attention: decomposed soft tissue is difficult to segment, and its structural variability challenges standard morphometric methods.

Studying this question requires imaging data that span the full range of decomposition. The New Mexico Decedent Image Database \citep[NMDID;][]{edgar2020new} provides whole-body computed tomography (CT) scans and metadata for more than 15{,}000 decedents to approved researchers. Previous applications include studies of age at death and sex using skeletal and dental structures \citep{ruff2024effects,joshi2023three,kuchavr2024automatic, appel2023error}, but decomposition of internal soft organs has received limited attention. We use NMDID to analyze brains and kidneys from 59 decedents spanning four whole-body decomposition stages. To our knowledge, this is \textit{the first organ-level statistical shape analysis of postmortem decomposition in internal soft tissue}.

Decomposed organs present an unusual shape-analysis problem in which both geometry and topology vary: internal cavities develop, organs shrink, and tissue can fragment into disconnected pieces \citep{vass2001beyond,roberts2012post}. Because internal voids are among the most informative signatures of decay, outer-surface representations are insufficient; each organ must be represented using both its outer surface and its internal cavity boundaries. At the same time, most classical shape analysis methods presume landmark or point correspondences that identify the same anatomical locations across observations \citep{dryden2016statistical}, typically obtained by manual or automated registration \citep[e.g.][]{besl1992method}. This presumption is reasonable when organs share a common structure and differ only moderately, but it breaks down for decomposed organs because structural integrity varies widely. When one kidney is intact and another has internal voids or only a few fragments, correspondence is hard to estimate and arguably undefined. Registration can still be forced, but the resulting alignments introduce distortions that propagate into subsequent analysis.

Topological data analysis (TDA) offers a way to compare these structures without anatomical correspondence or explicit registration \citep{edelsbrunner2002topological,wasserman2018topological}. Persistent homology computed from a distance-based filtration treats each point cloud as a self-contained object. It is invariant to rigid motions, requires no alignment, and directly captures features that decomposition creates: connected components, loops, and enclosed voids. TDA has an established record in biomedical imaging applications, including shape analyses of brain artery trees \citep{bendich2016persistent}, tumor imaging \citep{crawford2020predicting, moon2023using}, and brain network analyses of PET and MRI data \citep{wu2023topological, rajakaruna2026enhancing}; for point clouds, elastic alignment of persistence landscapes separates topological signal from geometric and sampling variation \citep{matuk2024topo}. Other correspondence-free approaches have different orientation and computational requirements. Directional persistent homology and Euler characteristic transforms \citep{turner2014persistent, meng2025randomness} require accounting for orientation when comparing shapes, while Gromov--Wasserstein distances \citep{memoli2011gromov} are isometry-invariant but computationally demanding for large point clouds. For our data, persistent homology of the alpha filtration is feasible for point clouds containing hundreds of thousands of points.

In this study, we analyze brains and kidneys from 59 decedents across four decomposition stages, representing each organ as a boundary point cloud. We first evaluate a registration-based analysis using iterative closest point (ICP, \citealp{besl1992method}) alignment to a template, followed by principal component analysis and a fast approximation of principal nested cones \citep{zhan2026principal}. This pipeline yields an interpretable ordering for brain decays but less coherent results for kidneys, whose decay patterns are more heterogeneous. We then develop a registration-free pipeline based on persistent homology: complementary topological summaries and distances are fused, with the combination selected in a simulation study of progressive decay, and the fused distances support clustering and seriation. The pipeline yields interpretable organ-specific \textit{decomposition atlases}, ordered reference collections of the observed morphologies, from intact to severely decomposed. Atlas rankings agree only partially across organs and with whole-body stages: agreement is greater between paired kidneys than between brains and kidneys, and whole-body stages correlate more strongly with brain rankings than with kidney rankings. These results suggest that decomposition can be asynchronous across organs within a body and provide organ-resolved support for a long-standing concern in forensic science that external whole-body scores only partially reflect internal organ-level state \citep{zhou2011factors}.

Another contribution is the data itself. Upon publication, we will publicly release the organ point clouds, whole-body stages, and analysis code (Section~\ref{sec:datarelease}). Each case represents up to five working days of expert CT segmentation \citep{tiago2026taphonomic}, and to our knowledge, no comparable public collection includes shape data for decomposed organs. The collection provides three-dimensional shapes with \textit{unreliable point correspondence} for developing and comparing statistical methods for shape analysis, topological inference, and ordering.
 

\section{Data and preprocessing}
\label{sec:data}

\subsection{NMDID dataset and study cohort}
\label{sec:cohort}

This study uses CT scans from NMDID~\citep{edgar2020new}, which includes more than 15,000 individuals who died in New Mexico between 2010 and 2017. Each individual is represented by approximately 10,000 DICOM slices, acquired at 1~mm thickness with 0.5~mm overlap. The accompanying metadata include about 60 metadata variables describing demographics, medical history, and circumstances of death. Access is available to approved application.

Among the metadata, NMDID includes a decomposition score assigned by CT technologists at the time of scanning according to the four stages of postmortem change described by \citet{lee2009early}: no decomposition, mild decomposition, moderate decomposition, or advanced decomposition, hereafter referred to as Stages~I--IV. The stage score assesses the \textit{whole body} rather than any particular organ, and it is an operational judgment made for scanning, not a validated forensic measure. \citet[Ch.4]{tiago2026taphonomic} also identified subjects whose organ morphology appeared inconsistent with their assigned stage. We therefore treat the decomposition-score stage as useful but approximate context rather than organ-level ground truth, which leads us to an unsupervised, organ-level analysis.

The study cohort was assembled as described below, with full details provided by \citet{tiago2026taphonomic}. We first drew 120 cases from NMDID, 30 per stage with available demographics, manner of death, and decomposition scores. To limit medical and toxicological confounding, we then excluded individuals with indications of medical treatment, such as chemotherapy, drug exposure, and alcohol use, leaving 59 natural-death cases (13 in Stage~I, 13 in Stage~II, 18 in Stage~III, and 15 in Stage~IV). The cohort size is capped by the cost of processing, as organ segmentation requires up to five working days per case. 

Beyond the decomposition score, sex, age, city of death, and the time delay between death and the CT scan are relatively complete records among the 59 decedents. Figure~\ref{fig:age_delay_stage} shows the distributions of selected metadata variables across whole-body decomposition scores. In a proportional-odds cumulative logit model for decomposition stage, drop-one term likelihood-ratio tests showed no evidence of associations with age (\(p=0.89\)) or sex (\(p=0.38\)), after adjustment for the other predictors. The time delay between death and the CT scan, however, was negatively associated with the decomposition stage (\(p=0.005\)), indicating that more decomposed bodies have a shorter time between cadaver collection and scan. The explanation is operational. 
NMDID times the delay from the declaration of death by an investigator rather than from biological death. Therefore, the time delay after death relates to when the body was discovered and the delay until scanning. Bodies in advanced decomposition tend to be transferred and scanned promptly once found, whereas hospital deaths wait for transport due to an administrative lag. The delay is therefore not a postmortem interval.

\subsection{Segmentation and point-cloud representation}

Brains and kidneys were selected for segmentation, surface extraction, and three-dimensional reconstruction because their contrast with surrounding tissues enabled clear distinction on CT \citep[Ch. 4.3,][]{tiago2026taphonomic}. The two organs also complement each other well for an organ-level study of decomposition. The brain is enclosed by the skull, which provides a persistent external boundary even at advanced decay, while the paired kidneys enable within-subject comparison of two anatomically similar organs. The two organs also differ in tissue decomposition because of their surrounding anatomical environment, making their comparative decomposition patterns biologically meaningful.

One operator segmented each organ semi-automatically, slice by slice, in Amira 5.4.5. Voxels within the regions of interest were delineated on each slice to define the target structures. The delineated segments were then stacked to reconstruct three-dimensional models. Intra-operator consistency was confirmed by repeating segmentations several months later \citep[Ch.~4.4,][]{tiago2026taphonomic}.

Each reconstructed organ is represented by a point cloud sampled from the boundaries of the segmented tissue, including the outer surface and the walls of any internal cavities. Coordinates are in millimeters in the scanner frame. This volumetric boundary representation is a key difference between studying decomposed and living organs: for living organs, a surface point cloud is usually sufficient, whereas decomposed organs demand a representation that can express internal cavities and fragmentation.

The reconstructed point clouds vary widely in size: point counts range from 31,723 to 280,664 for brains and from 1,667 to 31,595 for kidneys (Fig.~\ref{fig:ss}). Point count alone is a coarse but non-monotonic indicator of decay, as intact organs are represented mainly by their outer surface and contain relatively fewer points; organs with intricate internal cavities contain many points; and severely decayed organs can again contain very few points when only fragments remain.

\subsection{Observed morphological variability}
\label{sec:morphvar}

\begin{figure}[ht!]
    \centering
    \includegraphics[width=1 
                    \linewidth]{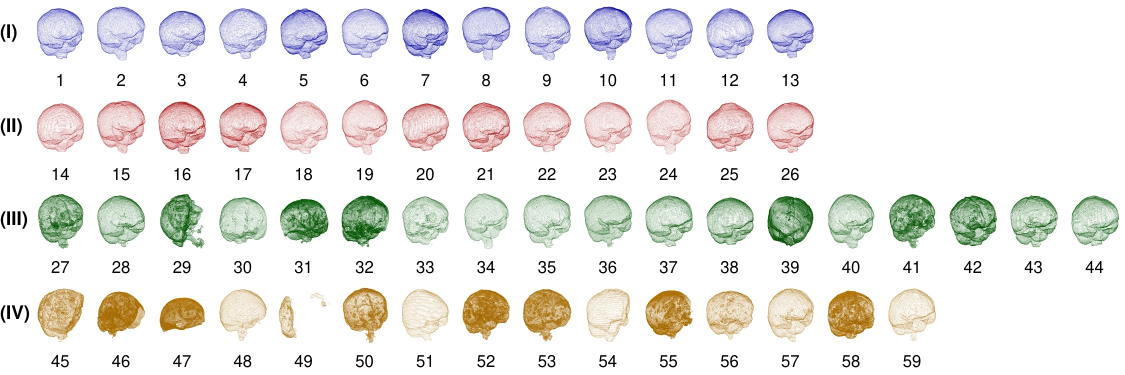}
    \caption{Reconstructed brains grouped by whole-body stages: (I)--(IV) denote Stages~I--IV. Samples are ordered by research IDs within each row. Brain~\(i\) corresponds to Specimen~\(i\). Interior points indicate internal voids.}
    \label{fig:br_3d_orig}
\end{figure}

\begin{figure}[ht!]
    \centering    
    \includegraphics[width=1 
                  \linewidth]{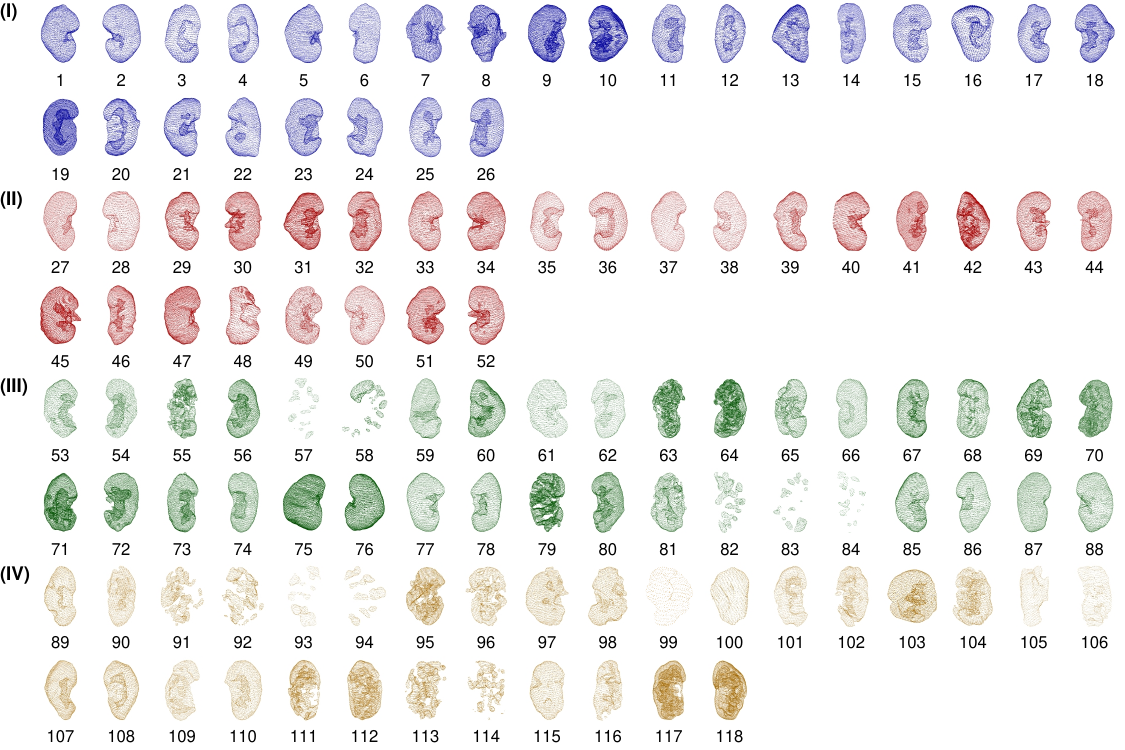}
    \caption{Reconstructed kidneys, arranged as in Fig.~\ref{fig:br_3d_orig}. For Specimen~\(i\), the left and right kidneys are indexed as \(2i-1\) and \(2i\), respectively.}
    \label{fig:kd_3d_orig}
\end{figure}

The complete collections of organs are shown in Figs.~\ref{fig:br_3d_orig} and~\ref{fig:kd_3d_orig}, grouped by stage. Specimens are numbered as 1--59, ordered by their research IDs. Specimen~$i$ contributes Brain~$i$ and Kidneys~$2i-1$ (left) and~$2i$ (right).
To facilitate visualization, the original spatial arrangement of the kidney pairs has been removed. Three preliminary observations motivate the remainder of the study:

\begin{itemize}
\item Brains and kidneys decay differently. Brains show two decay patterns: formation of internal voids alone (e.g., Brains~27 and~41), and degradation of the external structure together with formation of internal voids (e.g., Brains~29 and~47). 
Kidney decomposition is more varied. Some kidneys decay from the hilum, where the renal vein, artery, and ureter enter, producing a large unilateral cavity (e.g., Kidney~3); some become porous from the inside outward, forming many small internal holes (e.g., Kidney~64); some shrink in overall shape without developing clear cavities (e.g., Kidney~105); and some fragment into only a few disconnected pieces (e.g., Kidney~57).

\item The whole-body decomposition stage labels and the observed organ morphology can disagree, particularly in Stages~III and~IV. Some organs are labeled as moderately or severely decomposed, yet appear visually intact, such as Brains~51 and~59 and Kidneys~87 and~115. On the other hand, some are labeled as moderately decomposed yet show advanced destruction, such as Brains~29 and~41 and Kidneys~57 and~82. 

\item The brain and kidneys from the same body can show markedly different levels of decomposition. For example, Brain~40 in Fig.~\ref{fig:br_3d_orig} is nearly intact while the same specimen's kidneys (Kidneys~79 and~80 in Fig.~\ref{fig:kd_3d_orig}) show more advanced decomposition. Even the paired kidneys of one body can differ, e.g., Kidneys~79 vs.~80 and Kidneys~81 vs.~82.
\end{itemize}

These observed discrepancies motivated an organ-level unsupervised analysis in which brains, left kidneys, and right kidneys are treated as separate entities. The primary analysis was based on point-cloud representations themselves, with stage labels retained only as whole-body context. 
The aims are: (i) to cluster each organ collection into data-driven decomposition groups and (ii) to construct an organ-specific decomposition atlas by ordering the degree of decay for each organ collection. 


\subsection{Released data products}
\label{sec:datarelease}

To support reuse, we will deposit the processed point-cloud data and analysis code underlying this study in a public \href{https://github.com/YanyanZhan/shape-topological-brains-kidneys}{GitHub repository} upon publication. The release will include 177 organ point clouds (59 brains and 118 kidneys), with metadata on organ type, kidney laterality where applicable, and NMDID whole-body decomposition stage.

\section{Registration-based geometric analysis}
\label{sec:pcapnc}

A natural analysis approach is to stack the point coordinates of each registered organ into a single long vector and apply multivariate methods after registering the points so that they correspond in some meaningful way. We used principal component analysis \citep[PCA;][]{hotelling1933analysis,
jolliffe02} and a fast approximation of principal nested cones \citep[fast PNC;][]{zhan2026principal}. PNC is a nonlinear, backward-style dimension reduction method which successively projects observations onto nested lower-dimensional cones, and follows earlier work on principal nested spheres \citep{Jungetal12}. Unlike PCA, PNC retains size explicitly with a clear geometric interpretation, which carries meaningful information here, since organ size changes during decomposition. Fast PNC first applies a PCA-based transformation and then fits standard PNC in the reduced space, making it feasible for high-dimensional point clouds.

\subsection{Point cloud registration}
\label{sec:regist}

Both PCA and fast PNC require compatible representations of all samples, with the same number of points in corresponding order. We aligned each sample to a common template using ICP in {\tt geomorph} \citep{AdamsCollyer2026}, with right kidneys reflected beforehand to match the orientation of left kidneys. Point-wise correspondences were then taken as the nearest neighbors to the template. Rigid ICP was used to estimate translations and rotations and scaling was not permitted.
The results presented in this section were obtained using templates constructed from Brain~1 and Kidney~1.
Since interior cavities represent important decay features, each template contains both interior and surface points (Fig.~\ref{fig:ICPtemp}).
Template construction details can be found in Supplementary Materials Section~\ref{sec:Template}.
Brain~49 survives only as a small fragment that cannot be meaningfully matched to the template and distorts the fitted components of both methods. We therefore excluded Brain~49 from the registration-based analyses here, although no exclusion was necessary for the registration-free analysis in Section~\ref{sec:tda}. 

For each object type after carrying out ICP we have a registered, centered sample of $k \times 3$ matrices $\boldsymbol{x}_i, i=1,\ldots,n$. PCA is carried out using the sample covariance matrix of these data. 
The PNC method fits cones of successively lower dimensions with opening angles obtained by minimizing the sum of squared radial residuals, and slant heights given by the Frobenius norm $\|\boldsymbol{x}_i\|_F, i=1,\ldots,n$.

\subsection{PCA and fast PNC results}
\label{sec:pca_pnc_res}

\begin{figure}[ht!]
    \centering 
    \includegraphics[width=1 
            \linewidth]{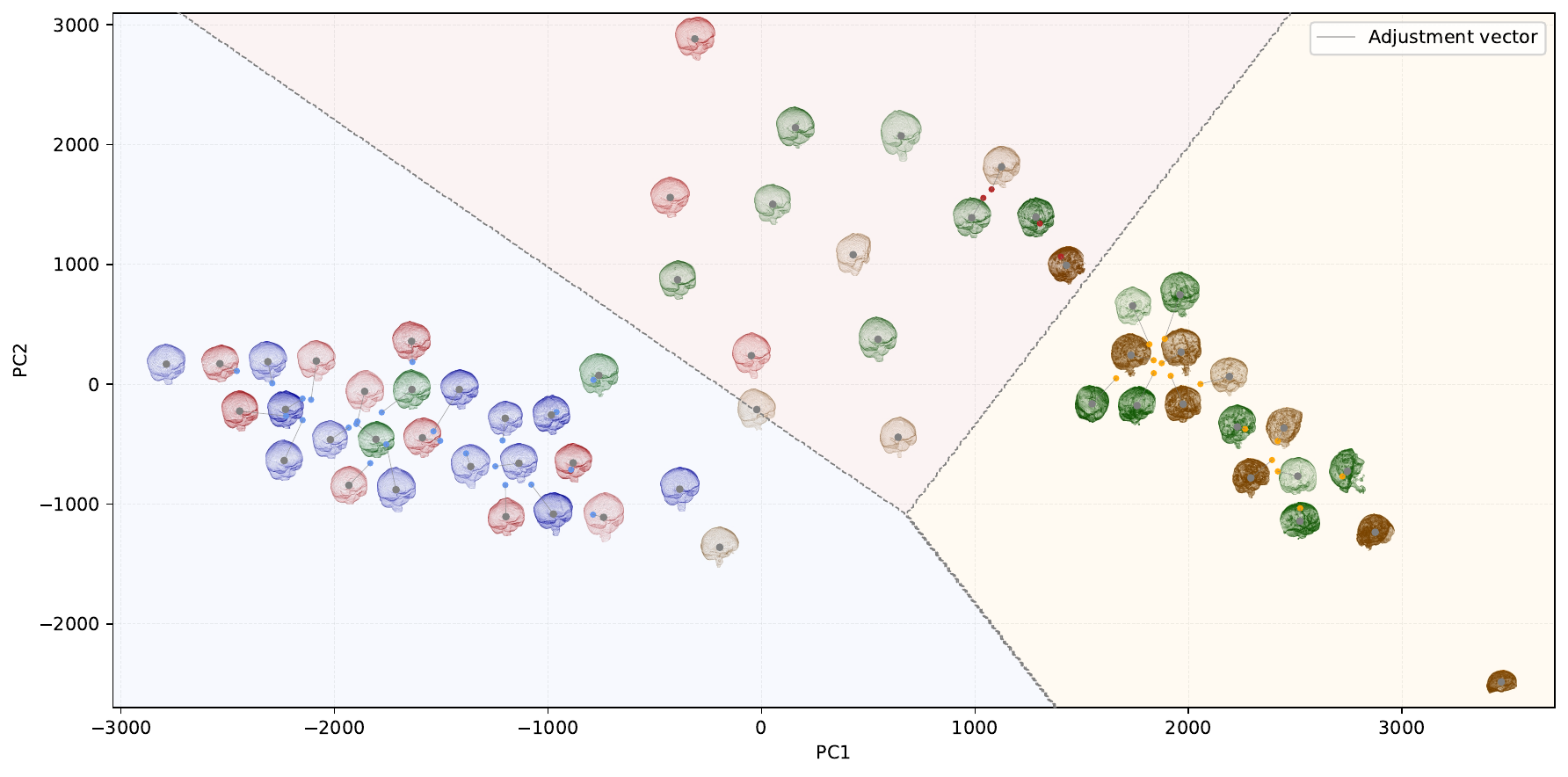}
    \includegraphics[width=1  
             \linewidth]{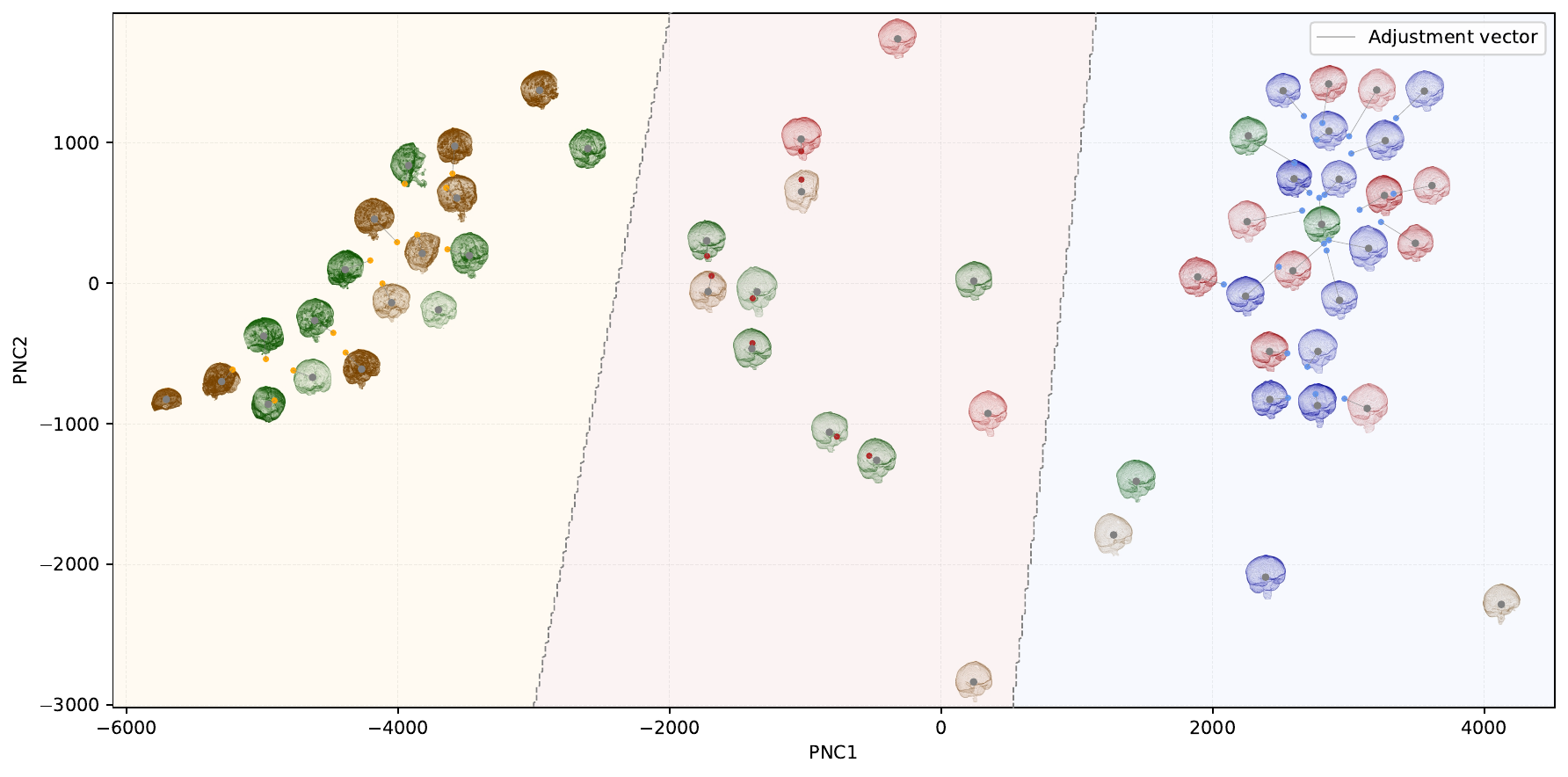}
    \vspace{-5pt}
    \caption{Scores of the brains on the first two components from PCA (top; PC1 and PC2 explain 32.2\% and 9.9\% of the variation) and fast PNC (bottom; PNC1 and PNC2 explain 59.7\% and 5.1\%), colored by whole-body stage. Overlapping thumbnails are shifted slightly for display; gray markers show adjusted positions. Shaded regions show linear discriminant analysis boundaries fitted to $k$-medoids clusters($k=3$; Table~\ref{tab:pnc}), for display only.}
    \label{fig:pca_embed_br}
\end{figure}

\begin{figure}[ht!]
    \centering 
    \includegraphics[width=1 
        \linewidth]{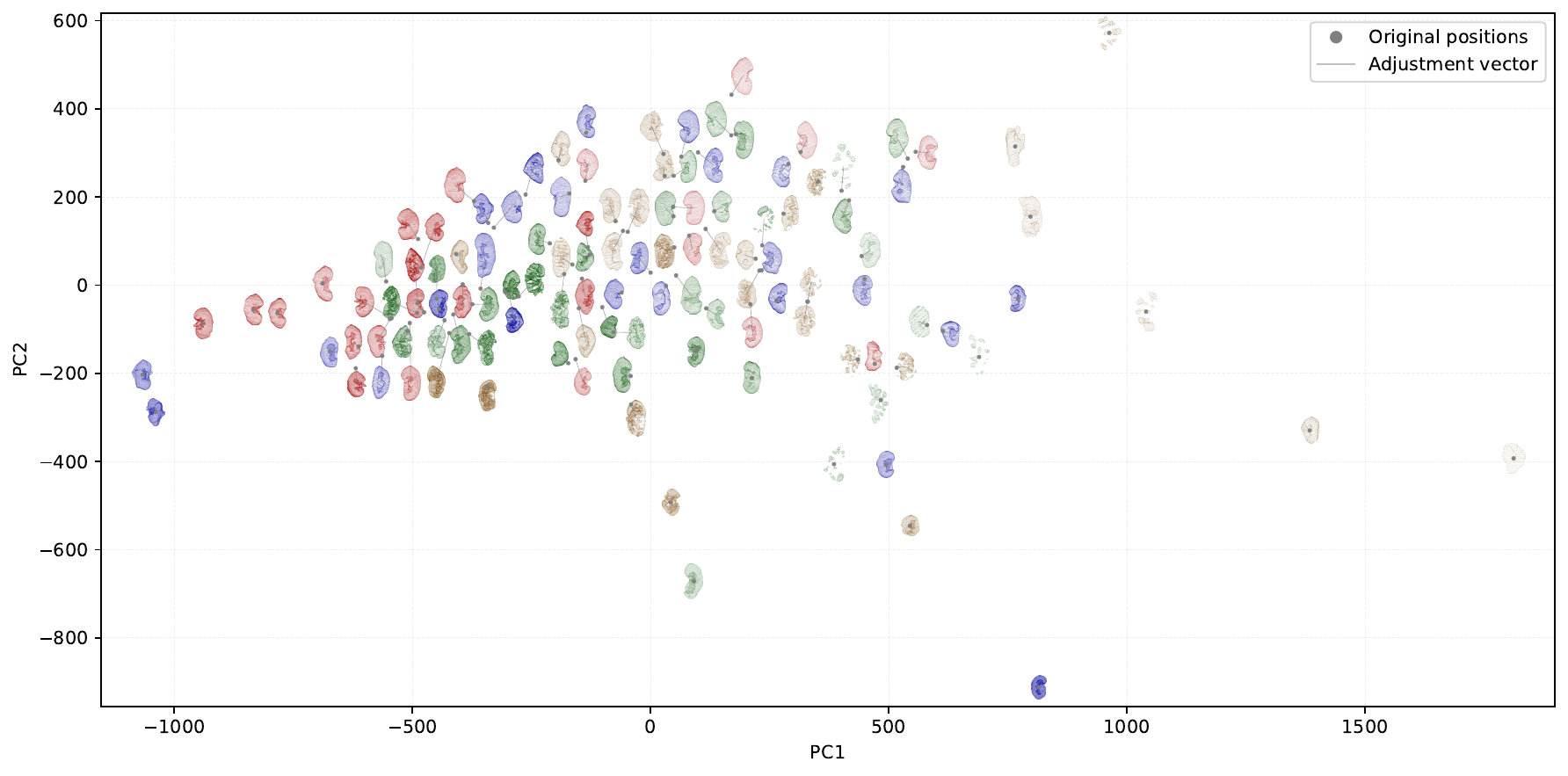}
    \includegraphics[width=1  
         \linewidth]{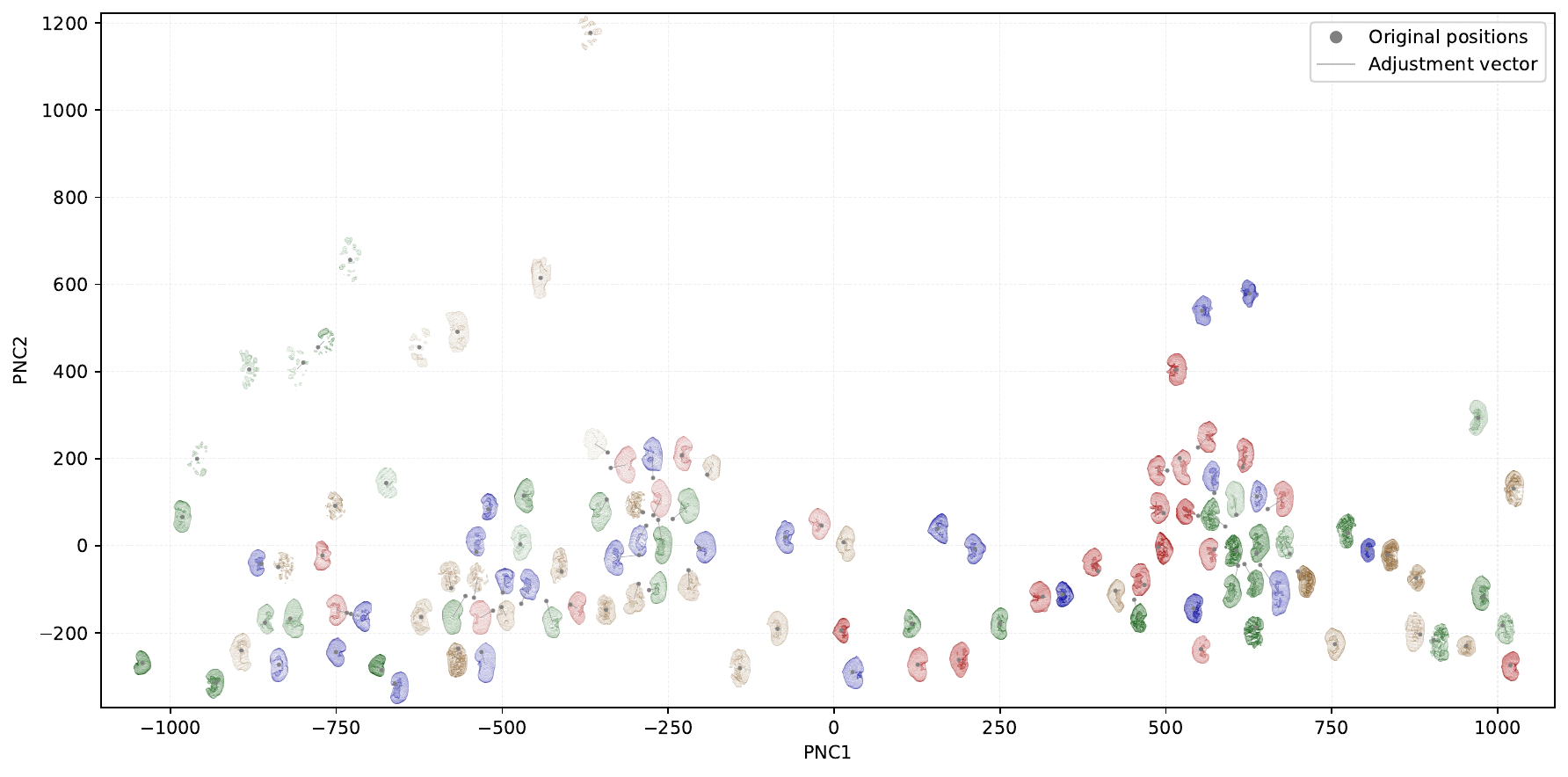}
    \vspace{-5pt}
    \caption{Scores of the kidneys on the first two components from PCA (top; PC1 and PC2 explain 29.9\% and 6.3\% of the variation) and fast PNC (bottom; PNC1 and PNC2 explain 36.5\% and 5.9\%), colored by whole-body stage. Overlapping thumbnails are shifted slightly for display; gray markers show adjusted positions. No clustering is shown because neither score plot exhibits a clear grouping of the samples.
    }
    \label{fig:pca_embed_kd}
\end{figure}

For brains, both PCA and fast PNC produce reasonably interpretable results. As shown in Fig.~\ref{fig:pca_embed_br}, both PC1 and PNC1 broadly reflect the progression of brain decomposition, with higher PC1 values and lower PNC1 values corresponding to more severe decay. Figure~\ref{fig:br_pnc_polar} shows the brains embedded by their fast PNC polar scores, which form a circular pattern ordered by decomposition severity.
In contrast, neither method yields meaningful ordering or separation for kidneys. As shown in Fig.~\ref{fig:pca_embed_kd}, samples with different decomposition status are heavily intermixed, preventing an interpretable representation of kidney decomposition.

This contrast is consistent with differences in registration quality. Brain decay is sufficiently stereotyped and retains enough common structure for alignment to a single template, providing meaningful correspondences for downstream analysis. In contrast, kidney decay shows greater heterogeneity, which complicates registration. When only a small portion of an organ remains, establishing meaningful correspondences becomes difficult, and in some cases, it is unclear whether any alignment is sensible. As a result, registration-dependent methods are not optimal.

Nearest-neighbor registration in ICP adds an additional artifact. Point counts vary widely across samples (Fig.~\ref{fig:ss}), and enforcing equal point counts can introduce local duplication. For samples containing only a small number of voids, the void points become nearest neighbors of many interior template points and are over-represented. This artifact may explain the sharp separation between brains with a few cavities and those without cavities in Fig.~\ref{fig:pca_embed_br}. Correspondence ambiguity and repeated representation motivate the registration-free analysis presented in Section~\ref{sec:tda}.

To assess the sensitivity to the template choice in ICP, we repeated the registration and ranking with alternative templates for each organ, with details and results included in Section~\ref{sec:Template}. The brain orderings agree closely across templates ($\tau \ge 0.79$), so the geometric ordering used in Section~\ref{sec:atlas} does not depend greatly on the particular template. Agreement across kidney templates is much lower (PNC $\tau$ as low as 0.43), indicating no common structure supports correspondence; the kidney registration inherits its reference, and so does any ordering built on it. This is one of the reasons to construct the kidney atlas from the registration-free analysis alone in Section~\ref{sec:atlas}.

\section{Topological data analysis}
\label{sec:tda}
Topological data analysis (TDA) describes a shape through its connectivity structure, including connected components, loops, enclosed voids, and the scales at which these features appear and disappear. TDA processes each point cloud independently, so no cross-sample registration is needed. For distance-based filtration, the resulting summaries are invariant to translation, rotation, and reflection, so the samples need no rigid transformation and Brain~49 poses no additional difficulty. 

Our pipeline consists of three steps: computing a persistence diagram for each organ; summarizing the diagrams and calculating pairwise dissimilarities between organs; and fusing complementary dissimilarity measures into a single matrix for clustering and seriation. We select the combination of dissimilarity measures through a mechanistic simulation study (Section~\ref{sec:simu}) and then apply the fixed pipeline to the real organ data (Section~\ref{sec:tda_real_data_res}). Sections~\ref{sec:ph} and \ref{sec:dist} review only the TDA concepts used in this study. For more comprehensive treatments,  see \citet{edelsbrunner2010computational} and \citet{wasserman2018topological}.

\subsection{Persistent homology}
\label{sec:ph}

Persistent homology \citep{edelsbrunner2002topological} tracks the topological features of a point cloud across scales. It is typically computed from a filtration, which is a nested sequence of complexes 
\(
    K_{0} \subseteq K_{1} \subseteq \cdots \subseteq K_{m}
\)
constructed by gradually increasing a scale parameter $\alpha$, which can be pictured as thickening each point into a ball of radius $\alpha$. To accommodate the large size of point clouds, we use the alpha-complex filtration \citep{edelsbrunner1994three,cazals2006delaunay}, a subcomplex of the Delaunay complex that is computationally efficient and carries the same topology as the union of balls at every scale. We report $\alpha$ as a ball radius in millimeters, which is the square root of GUDHI's \citep{maria2014gudhi} filtration value, so each scale corresponds directly to a physical distance.

As the filtration parameter $\alpha$ increases, the evolution of topological features is tracked across scales. Each feature is born at a scale $b$ and dies at a scale $d$. Collecting these birth--death pairs gives a persistence diagram,
\(
    D = \{(b_j,d_j)\}_{j=1}^J,
\)
where \(b_j \leq d_j\). The persistence diagram is a set of points with multiplicity in the upper half-plane. The lifetime $d_j - b_j$ separates stable structures from sampling noise: long-lived features reflect geometry, whereas short-lived features may reflect noise, sampling variability, or finer-scale local structures.

In $\mathbb{R}^3$, homology is nontrivial in dimensions \( q = 0 \) (connected components), \( q = 1 \) (loops), and \( q = 2 \) (enclosed voids). Since organ decay mainly produces internal voids, features in dimension \( q = 2 \) are of primary interest in this study. Before fixing this choice, we also constructed orderings from $H_0$ and $H_1$ (Section~\ref{sec:Dimension}, Figs. \ref{fig:h0_brain}–\ref{fig:h2_kidney}). The orderings induced by $H_0$ and $H_1$ do not track the progression of decomposition. For example, several heavily fragmented kidneys are placed at the intact end of the sequence, and $H_1$ places the most severely decayed brain near the start of the intact end. The lifetimes of $H_0$ and $H_1$ features are determined mainly by the sampling density of boundary-sampled point clouds. Thus, they mainly reflect processing differences, rather than decomposition status and organ structure. The forensic expert on the team also considered only $ H_2$ orderings anatomically plausible. Accordingly, all subsequent analyses focus on persistent homology \( H_2 \).

\begin{figure}[ht!]
    \centering
    \includegraphics[width=1 
            \linewidth]{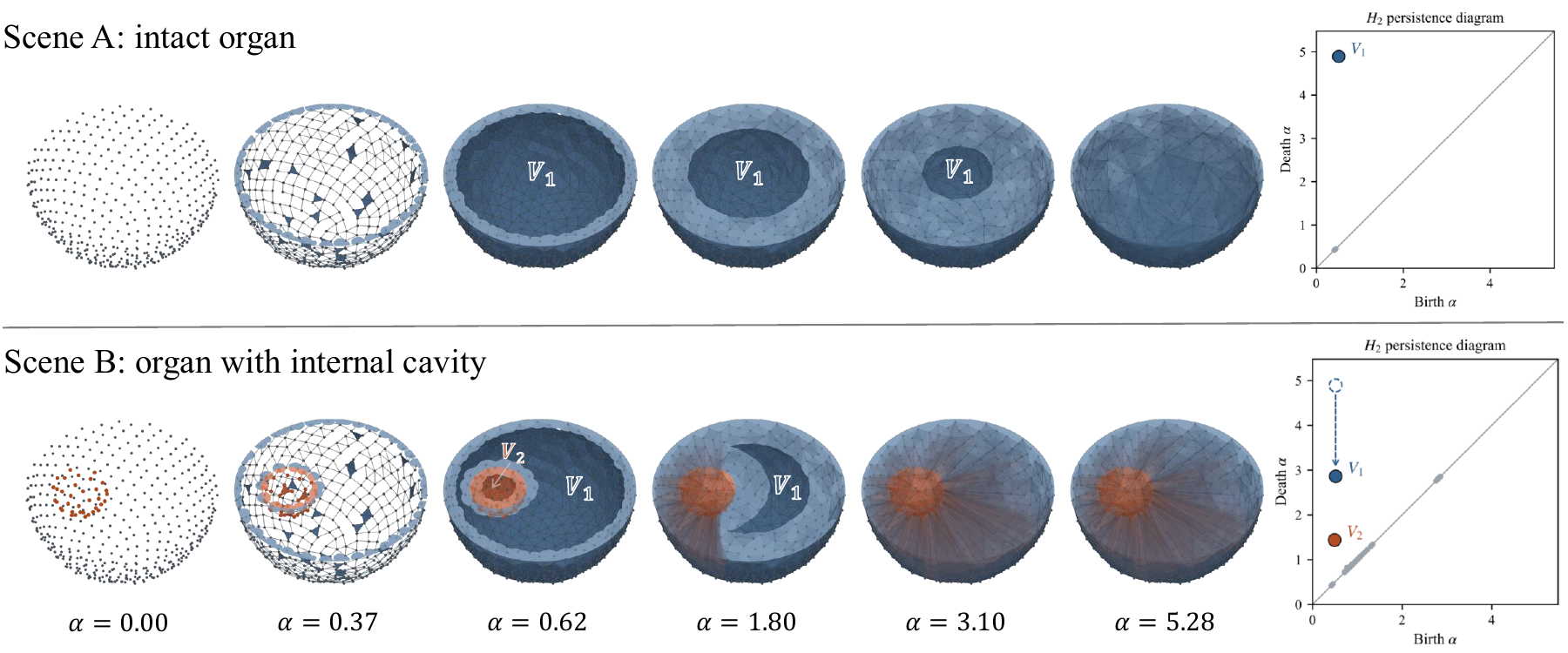}
    \caption{Alpha-complex filtration and $H_2$ persistence diagrams for Scene A (points on a sphere mimicking an intact organ) and Scene B (the same sphere plus an interior sphere mimicking an internal cavity boundary; orange points). The front half of each complex is removed to reveal points, edges, and triangles behind the cut. The semitransparent cut plane shows the cross-section of the union of balls, shaded orange within the original cavity, with underlying simplices faintly visible. Voids appear as open areas of the section, labeled as $V_1$ and $V_2$ while they persist. Gray points near the diagrams’ diagonals correspond to sampling noise.}
    
    \label{fig:tda-toy}
\end{figure}

For illustration, Fig.~\ref{fig:tda-toy} shows a 3D example that mimics the two organ conditions. In Scene A (top row), points are sampled approximately uniformly on a sphere of radius 5, representing the outer surface of an intact organ. In Scene B (bottom row), the same outer sphere is combined with a smaller interior sphere of radius 1.5, representing the boundary of an internal decomposition cavity. The leftmost panels show the two point clouds, and the remaining panels display the alpha complex as $\alpha$ increases. On the cut plane, the cross-section of the union of balls of radius $\alpha$ is drawn; the alpha complex carries the same topology as this union of balls at every scale \citep{edelsbrunner2010computational}, so voids appear as open areas of the section and are filled in exactly at their death values. At small $\alpha$, the points first connect into edges and scattered triangles ($\alpha = 0.37$). Once $\alpha$ reaches the sampling scale, each sampled surface encloses a void: the interior void $V_1$ in both scenes, and additionally the cavity void $V_2$ in Scene B ($\alpha = 0.62$). As $\alpha$ increases further, the thickening grows inward, and each void is eventually filled. In Scene A, $V_1$ dies at $\alpha \approx 4.9$, close to the sphere radius. In Scene B, $V_2$ dies at $\alpha \approx 1.4$, close to the cavity radius, while $V_1$ now dies at $\alpha \approx 2.9$, because the cavity blocks the largest empty ball inside the organ. A cavity therefore leaves two traces: a new $H_2$ point and a shorter-lived outer void.


Persistence diagrams may contain both meaningful topological features and noise. Even in the simple example of Fig.~\ref{fig:tda-toy}, spurious features appear as short-lived points close to the diagonal. To reduce noise, we discard features with persistence below \(p_t\), the 90th percentile of all persistence values pooled across organs of one type. Pooling persistence at a single physical scale across all samples within an organ ensures the retained features are comparable. 
We ran a sensitivity analysis of percentile choice by rerunning the full pipeline with thresholds from the 85th to 95th percentile. Final orderings are nearly identical for brains and stable for kidneys up to the 91st percentile, with all pairwise Kendall's $\tau$ at least 0.94. Details are given in Section~\ref{sec:Denoising}. 

\subsection{Topological summaries and distances}
\label{sec:dist}
Persistence diagrams are multisets of varying numbers of points, so we compare them either directly using matching-based metrics or through four fixed-form summaries, which capture different aspects of the diagrams. 

Persistence entropy \citep{chintakunta2015entropy} is a scalar measure that quantifies how feature lifetimes distribute. Based on Shannon entropy of the normalized lifetimes, 
\begin{equation*}
    \mathrm{PE}(D) = - \sum_{j=1}^{J}p_j \log (p_j) \quad \text{and} \quad p_j = \frac{d_j - b_j}{\sum_{j'} (d_{j'} - b_{j'})}.
\end{equation*}
It reflects the level of disorder in the diagram. $\mathrm{PE}(D)$ is high when many features have comparable lifetimes and low when one or a few prominent topological features exist.

The persistence landscape \citep{bubenik2015statistical} provides a functional summary to record feature profiles by scale. For each birth--death pair $(b_j,d_j)$, define the tent function
\begin{equation*}
    \Lambda_j(\alpha)=\max\{0,\min(\alpha-b_j,d_j-\alpha)\}.
\end{equation*}
The $k$th landscape layer is given by the $k$th largest tent value at each $\alpha$:
\begin{equation*}
    \lambda_k(\alpha)=\text{$k$th largest value of } \{\Lambda_j(\alpha)\}_{j=1}^J .
\end{equation*}
Thus, $\lambda_1$ records the largest persistence contribution at each scale $\alpha$, while higher-order landscape functions summarize less dominant features. In this paper, we consider the first three layers of landscapes, namely \( \lambda_1 \), \( \lambda_2 \), and \( \lambda_3 \).

The silhouette function \citep{chazal2014stochastic} combines the tents into a single functional summary through a weighted average,
\begin{equation*}
\phi^{(p)}(\alpha)
=
\frac{\sum_{j=1}^{J} |d_j-b_j|^p \Lambda_j(\alpha)}
{\sum_{j=1}^{J} |d_j-b_j|^p}.
\end{equation*}
We use $p=1$ in this paper, which favors persistent features without overweighting them.

The Betti curve $\beta(\alpha) =\sum_{j=1}^{J} \mathbf{1}_{[b_j,d_j)}(\alpha)$ records the number of topological features alive at each scale, describing how the topology evolves.

\begin{figure}[ht!]
    \centering
    \includegraphics[width=1 
          \linewidth]{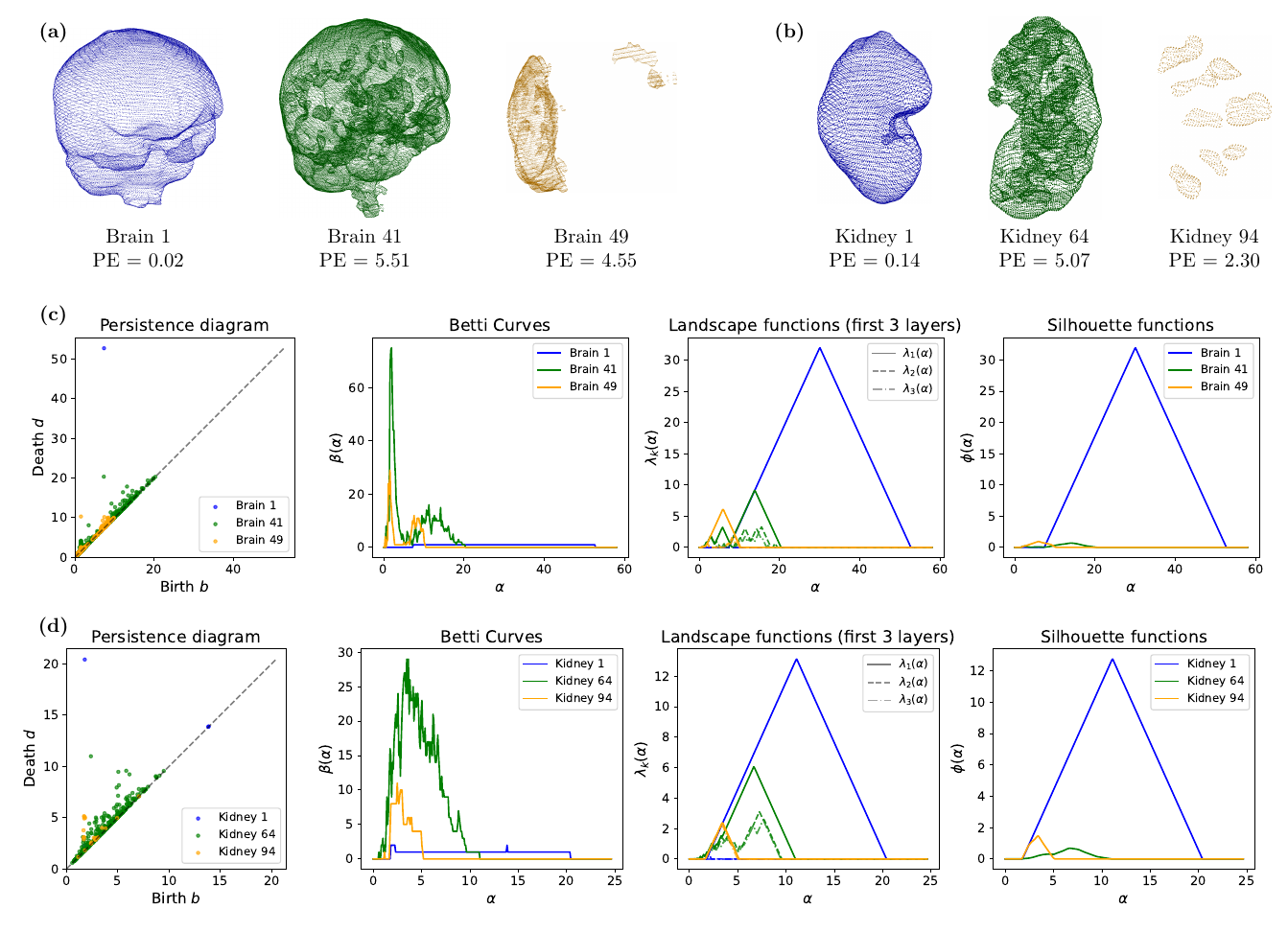}
    \vspace{-9pt}
    \caption{Example topological summaries for three brains and three kidneys. Panels (a) and (b) show the representative brain and kidney point clouds, with their corresponding IDs and entropies, colored as in Fig.~\ref{fig:br_3d_orig}. Panels (c) and (d) show the corresponding $H_2$ persistence diagrams, Betti curves, landscape layers $\lambda_1$--$\lambda_3$, and silhouettes. Scales are in mm; summaries are computed from the denoised diagrams.
    }
    \label{fig:feature-ex}
\end{figure}

Figure~\ref{fig:feature-ex} shows the persistence diagrams and their computed summaries for three representative brain samples and three kidney samples. For both organs, visually intact samples (Brain~1 and Kidney~1) have a single dominant feature, a near-zero $\mathrm{PE}$ value, and simple Betti curves, landscapes, and silhouette functions, suggesting a simpler topological structure. Organs with more advanced decompositions show more cavities (Brain~41 and Kidney~64) and have much higher \(\mathrm{PE}\) values. Their persistence diagrams contain more off-diagonal points, and their Betti curves show more topological features across a wider range of filtration scales. These patterns indicate increased structural complexity. The entropies of the most decayed samples (Brain~49 and Kidney~94) fall again because severe decomposition destroys the very structures that made the diagrams complex. Topological complexity is not monotone in decay, which is one reason no single summary suffices.

Diagrams can also be compared directly using diagram-based distances. Let $\Gamma$ be the set of matchings between two diagrams $D$ and $D'$: bijections between $D \cup \Delta$ and $D' \cup \Delta$, where $\Delta$ is the diagonal with infinite multiplicity, so any point may be matched to the diagonal rather than to a point of the other diagram. The bottleneck distance $d^{\mathrm{B}}$ is 
\begin{equation*}
  d^{\mathrm{B}}(D, D') = \inf_{\gamma \in \Gamma}\, \sup_{x}\, \lVert x - \gamma(x) \rVert_\infty,
\end{equation*}
which measures the largest discrepancy between corresponding topological features. The 2-Wasserstein distance $d^{\mathrm{W}}$ generalizes this idea by aggregating the costs of all matched point pairs between diagrams:
\begin{equation*}
  d^{\mathrm{W}}(D, D') = \left( \inf_{\gamma \in \Gamma} \sum_{x} \lVert x - \gamma(x) \rVert_\infty^{2} \right)^{1/2}.
\end{equation*}
Bottleneck distance $d^{\mathrm{B}}$ is coarser but stable under broad conditions, whereas $d^{\mathrm{W}}$ responds to every feature and is stable under additional conditions \citep{cohen2005stability,edelsbrunner2010computational,turner2014frechet}.

In addition to the two transport-based metrics, we used four \(L_2\) distances between the topological summaries: $d^{\mathrm{PE}}$ (absolute difference of entropies), $d^{\lambda}$ (on the concatenated layers $\lambda_1, \lambda_2, \lambda_3$), $d^{\phi}$ (on silhouette functions), and $d^{\beta}$ (on Betti curves). Table~\ref{tab:dists} lists the six distances, each of which captures a different aspect of persistence diagram dissimilarity.

\begin{table}[ht!]
    \centering
    \caption{Six distances between persistence diagrams. Each emphasizes a distinct aspect of dissimilarity.}
    \label{tab:dists}
    \begin{tabular}{llll}
        \hline
        & Distance & Representation compared & Primary emphasis \\
        \hline
        $d^{\mathrm{B}}$ & Bottleneck& diagrams (matching) & largest single-feature discrepancy \\
        $d^{\mathrm{W}}$ & 2-Wasserstein& diagrams (matching) & aggregate discrepancy over all features \\
        $d^{\lambda}$&Landscape & first three landscape layers, $\lambda_1$--$\lambda_3$ & dominant feature profiles across scales \\
        $d^{\phi}$&Silhouette & silhouette functions & persistence-weighted overall profile \\
        $d^{\beta}$&Betti & Betti curves & number of features alive across scales \\
        $d^{\mathrm{PE}}$& PE & persistence entropies & disorder of the lifetime distribution \\
        \hline
    \end{tabular}
\end{table}

For each point cloud, persistent homology was computed up to a maximal filtration value equal to the minimal extent of its bounding box. To make the functional summaries comparable across samples, all were then evaluated on a common grid of 300 evenly spaced points on the interval \([0, \alpha_{\max}]\), with $\alpha_{\max}$ fixed across samples within each organ collection: 57.98~mm for brains and 24.61~mm for kidneys. Summary-based distances provide a common framework for downstream analysis, although they are pseudometrics on the diagram space because different diagrams can share a summary.
Persistent homology was computed with GUDHI \citep{maria2014gudhi} on the full reconstructed point clouds. 

\subsection{Distance fusion, clustering, and seriation}

These summaries and distances provide complementary views of the same diagrams. At the same time, they are all computed from the same set of organ shapes, so they may capture overlapping structure. Combining information from multiple sources can therefore provide a more comprehensive profile of topological similarity and dissimilarity across samples. Before combining them, it is worth examining how much they actually agree, which we do with the Mantel statistic \citep{mantel1967detection}, the Pearson correlation between the corresponding off-diagonal entries of two dissimilarity matrices. Results are shown in Fig.~\ref{fig:compare}. All pairs are positively correlated, as expected for views of the same shapes. This shows that each feature captures part of the common structure in the data. However, the agreement between distances is far from uniform. They form two main groups, with $d^{\mathrm{PE}}$ and $d^{\beta}$ closer to $d^{\mathrm{W}}$, while $d^{\lambda}$ and
$d^{\phi}$ closer to $d^{\mathrm{B}}$, which indicates that different summaries emphasize different aspects of the persistence diagrams. The split is mild for brains and pronounced for kidneys, whose more varied decay routes are captured differently by different summaries.

This disagreement shapes how the distances should be combined. One approach is to analyze each distance matrix separately and then aggregate the outputs, using methods such as cluster ensembles \citep{strehl2002cluster} or rank aggregation \citep{dwork2001rank}. This works well when individual results broadly agree, but may be affected by outlying or inconsistent outputs. The kidney panel of Fig.~\ref{fig:compare}(b) shows exactly the inconsistency that makes the approach unsatisfactory here. We therefore combine the distances before analysis, in the spirit of similarity network fusion \citep{wang2014similarity}, except that our matrices are dissimilarities and the topological distances are fused directly. 

Let $d_m(i,j)$ denote the $m$th distance between organs $i$ and $j$, and let $s_m$ be the standard deviation of the off-diagonal entries of the $m$th matrix. To prevent a component from dominating solely because of its numerical scale, we define the fused distance as
\begin{equation}
\label{eq:fusion}
    d^{\mathrm{fus}}(i,j)
    =  
    \sqrt{
    \sum_{m=1}^{M}
    \left(
    \frac{d_{m}{(i,j)}}{s_m}
    \right)^2
    } \, .
\end{equation}
Here \(d^{\mathrm{fus}}(i,j)\) denotes the \((i,j)\) entry of the fused distance matrix \(\boldsymbol{d}^\mathrm{fus}\). If each component distance is pseudometric, then \(\boldsymbol{d}^\mathrm{fus}\) is also pseudometric by Minkowski's inequality. The remaining question is which component distances to include, and Section~\ref{sec:simu} provides an answer by simulation.

The resulting fused distance matrices can be used for distance-based analyses, such as clustering and ordering. We cluster with \(k\)-medoids, which directly uses pairwise dissimilarities, and order with seriation, which seeks a permutation \(\pi\) of \(\{1,\dots,n\}\) under which the permuted matrix is approximately banded, with smaller distances concentrated near the diagonal. We use the ARSA algorithm \citep{brusco2008heuristic} in the R package \texttt{seriation} \citep{hahsler2008getting}. A seriated ordering is defined up to reversal, we therefore orient it from relatively intact to more decomposed so that it correlates positively with whole-body stage labels.

\subsection{Mechanistic simulation studies}
\label{sec:simu}

Six candidate distances give 57 candidate fusions of two or more, and no existing theory indicates which individual distance or combination is more effective for recovering the decay ordering. We therefore compare them in a simulation of the dominant mechanism we observe: cavities appearing and growing inside an organ.

Briefly, each trajectory starts at \(t=1\) with points sampled uniformly on an ellipsoidal shell with semi-axes 8, 5.5, and 4. It accumulates internal spherical cavities over $T = 300$ time steps, two per step, with centers placed at spatially biased locations and radii growing linearly to a cap. The point cloud at time $t$ samples the boundary of the ellipsoid minus the union of its cavities. Full simulation details are provided in Section~\ref{sec:SimuDetails}. Figure~\ref{fig:simu_t} shows an example simulated trajectory at selected time points from a full sequence. 

From $N = 100$ simulated independent trajectories, we form 200 cross-sectional data sets. Each draws 50 trajectories with replacement and draws 50 out of the \(T=300\) time points without replacement and pairs them. Since the selected time points were known, they provide the ground-truth progression order for evaluation. In each set, we seriate with each of the six individual distances and, for each of the 57 subsets, two fusion methods: one based on Eq.~\eqref{eq:fusion} and the other using direct Borda aggregation of the separate rankings \citep{lansdowne1996applying}. Each of the $6 + 57\times 2=120$ orderings is compared by Kendall's $\tau$ with the true time order.

\begin{figure}[ht!]
    \centering    
    \includegraphics[width=1 
          \linewidth]{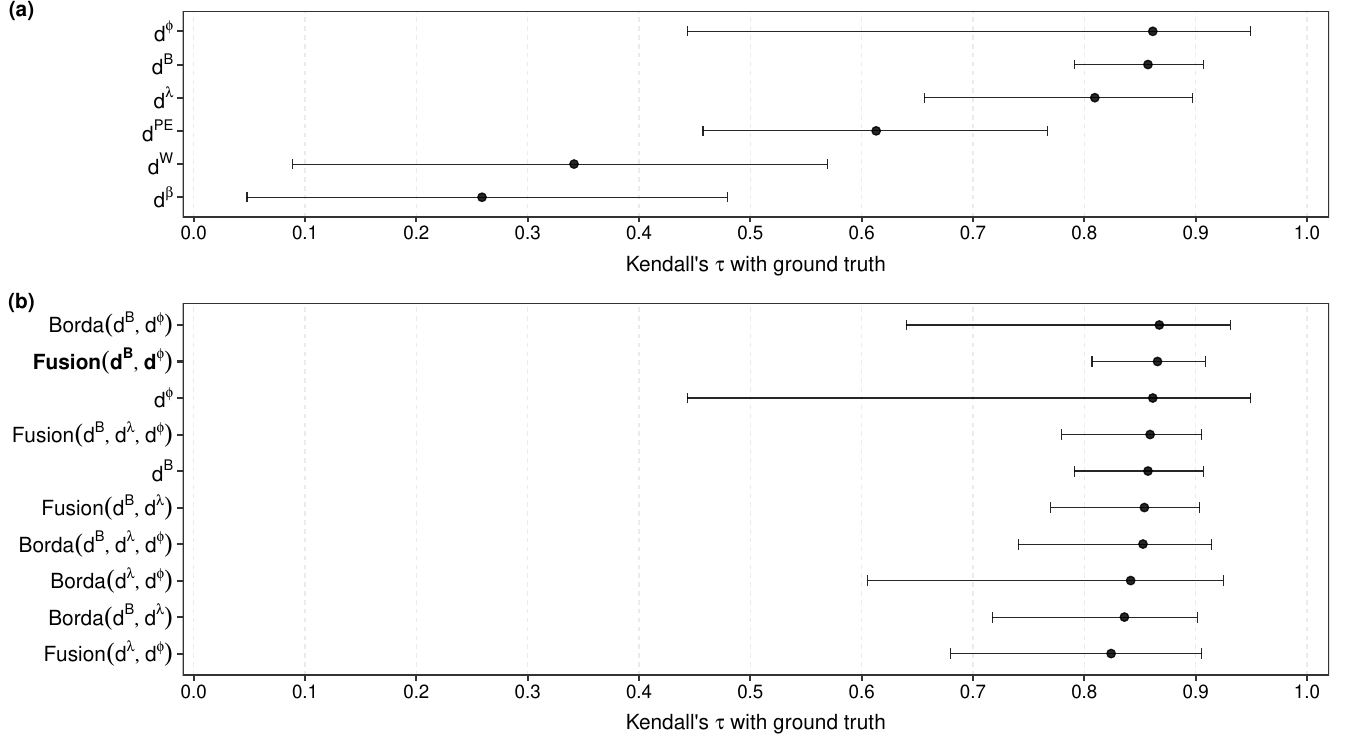}
    \caption{Mean Kendall's $\tau$ between the estimated and true orderings across 200 simulation repetitions, with error bars showing the 5th--95th percentile range. (a) Individual distances. (b) The ten methods with the highest mean $\tau$. Borda$(\cdot)$ aggregates the rankings computed separately from each listed distance using the Borda count; Fusion$(\cdot)$ combines the listed distances into a single distance via Eq.~\eqref{eq:fusion} before ranking.}
    \vspace{-10pt}
    \label{fig:simu_tau_ci}
\end{figure}

Among individual distances (Fig.~\ref{fig:simu_tau_ci}(a)), $d^{\mathrm{B}}$ has the best performance regarding both average accuracy and stability, while $d^{\mathrm{W}}$ and $d^{\beta}$ have the lowest accuracy. Figure~\ref{fig:simu_tau_ci}(b) shows the top 10 methods with the highest average Kendall's $\tau$ among all 120 methods. Fusing $d^{\mathrm{B}}$ with $d^{\phi}$ further improves on $d^{\mathrm{B}}$ alone in both accuracy and stability. Direct Borda rank aggregation of the same pair achieves a comparable mean but is less stable. We therefore fix $M = 2$ with components $d^{\mathrm{B}}$ and $d^{\phi}$ in Eq.~\eqref{eq:fusion}, and use this fused distance for all real-data analyses.
The pairing choice is also interpretable in terms of Table~\ref{tab:dists}: $d^{\mathrm{B}}$ tracks the dominant void, while $d^{\phi}$ summarizes the lifetime-weighted profile of all features. 

\subsection{TDA results for brains and kidneys}
\label{sec:tda_real_data_res}

For each observation, we constructed a filtration and computed a persistence diagram. We then denoised diagrams, obtained silhouette functions \(\phi(\alpha)\) and pairwise distance $d^{\mathrm{B}}$, and fused $d^{\phi}$ and $d^{\mathrm{B}}$ into $d^{\mathrm{fus}}$ for clustering and seriation.

Clustering was performed with $k$‑medoids. We chose the number of clusters $k$ using three internal indices~\citep{brock2008clvalid}: connectivity (lower is better), Dunn, and average silhouette width (higher is better). As shown in Table~\ref{tab:tda}, these indices do not always favor the same value of $k$; balancing them gave $k=2$ for brains and $k=3$ for kidneys. For kidneys, $k=3$ also yields interpretable morphological groups, ranging from largely intact kidneys, through those with cavities, to those that are fragmented or contain many voids.

The clusters differ in the extent and pattern of decay. Brain cluster~1 mainly consists of specimens in better condition, with no or very few interior voids, whereas brain cluster~2 contains more severely decayed specimens.
For the kidney, cluster~1 corresponds to lightly decayed organs; cluster 2 is characterized by a large unilateral cavity located near the hilum; and cluster~3 contains specimens with many small interior voids or disconnected fragments, matching the decay patterns described in Section~\ref{sec:morphvar}. In MDS embeddings of $d^{\mathrm{fus}}$ (Fig.~\ref{fig:tda_br_kd_mds}), both organs are arranged roughly by degree of decay along the first coordinate.

We summarized each cluster by an approximate Wasserstein barycenter of
persistence diagrams~\citep{turner2014frechet, lacombe2018large}. As shown in Fig.~\ref{fig:bary_br}, the barycenter of brain cluster~1 contains one prominent long-persistence void, compared with a halved dominant lifetime plus many short-lived voids for cluster~2. Across the three kidney clusters, the dominant lifetime decreases progressively while the number of small, local voids increases. These patterns are consistent with observations from Fig.~\ref{fig:tda_br_kd_mds}.

We use permutational multivariate analysis of variance (PERMANOVA; \citealp{anderson2001new}) to assess how much variation in the topological dissimilarity matrix is explained by each grouping (Table~\ref{tab:adj_r2}, adjusted \(R^2\)). The whole-body decomposition stage explains 46--54\% of the variation among brains but only 20--23\% among kidneys. The organ-specific \(k\)-medoids clusters explained more variation: 81--87\% for brains and 70--75\% for kidneys. Overall, whole-body stage aligns more closely with brain topology than with kidney, and does not fully capture organ-specific variation.

\begin{table}[ht!]
    \centering
    \caption{Adjusted $R^2$ from PERMANOVA applied to the topological distance matrices, using whole-body decomposition stages (I--IV) and organ-specific $k$-medoids clusters.}
    \label{tab:adj_r2}
    \begin{tabular}{c|ccc|ccc}
    \hline
       & \multicolumn{3}{c|}{Brains} & \multicolumn{3}{c}{Kidneys} \\
       Grouping  & $d^{\mathrm{B}}$ & $d^{\phi}$ & $d^{\mathrm{fus}}$ 
                 & $d^{\mathrm{B}}$ & $d^{\phi}$ & $d^{\mathrm{fus}}$ \\
    \hline
        Whole-body decay stages & 0.4621 & 0.5424 & 0.4901 & 0.2237 & 0.1963 & 0.2160 \\
        Organ-specific clusters & 0.8164 & 0.8741 & 0.8331 & 0.7471 & 0.6957 & 0.7162 \\
    \hline
    \end{tabular}
\end{table}

\section{Decomposition atlas}
\label{sec:atlas}
The organ-specific clusters in Section~\ref{sec:tda_real_data_res} provide a categorical overview of broad patterns of internal decay. To complement these views, we construct organ-specific decomposition atlases: ordered reference collections that arrange each organ type from relatively preserved to severely decomposed. 
Each atlas combines the PNC and TDA orderings when PNC yields interpretable results; otherwise, it relies on the TDA ordering alone.
Since the data are cross-sectional, the atlas represents morphological severity rather than a longitudinal decomposition trajectory. 

We assessed the sensitivity of the TDA orderings to Gaussian perturbations of the point-cloud coordinates. Agreement with the original ranks remained high at $\sigma = 0.5$~mm ( Section~\ref{sec:Stability}), with mean Kendall's $\tau$ of approximately 0.84 for brains and 0.94 for kidneys.

\subsection{Brain atlas}
The PNC-based and topological brain orderings broadly agree but differ for some samples, so we use rank aggregation to combine their complementary information. Although Brain~49 was excluded from the PNC fit, we assigned it the PNC rank corresponding to the most advanced decomposition before performing rank aggregation, reflecting its near-total tissue loss.

We compared several common rank aggregation methods, including Borda aggregation \citep{lansdowne1996applying}, approximate Kemeny aggregation \citep{dwork2001rank,ConsRank}, and robust rank aggregation \citep[RRA;][]{kolde2012robust,RobustRankAggreg}. The pairwise Kendall's $\tau$ across methods ranges from 0.93 to 0.98 (Fig.~\ref{fig:rank_agg_comp}), suggesting that different rank aggregation methods yield nearly identical results here. Considering its computational efficiency, deterministic nature, intuitive interpretation, and lack of tuning parameters, we use Borda aggregation. 

\begin{figure}[ht!]
    \centering     
    \includegraphics[width=0.65\linewidth]{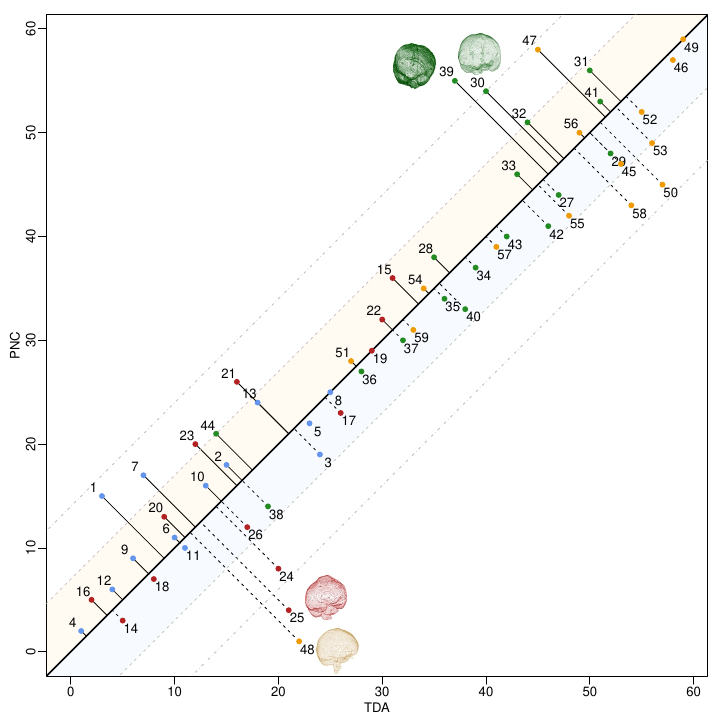}
    \vspace{-5pt}
    \caption{Borda aggregation of the two brain orderings. Each brain is placed by its TDA rank (horizontal) and PNC rank (vertical); its orthogonal projection onto the diagonal gives its Borda rank score. Points on the diagonal are ranked identically by the two methods; those above (below) are ranked later by PNC (TDA). The shaded band and the dot-dashed lines mark $\pm 1$ and $\pm 2$ standard deviations of the paired rank differences. Samples outside the $\pm 2$ SD band indicate substantial disagreement between the two ranks and are shown as thumbnails.}
    \label{fig:rank_br_comb}
\end{figure}

\begin{figure}[tbp]
  \centering
  \makebox[\linewidth][l]{\small (a)}\\[-12pt]
  \includegraphics[width=.7\linewidth]{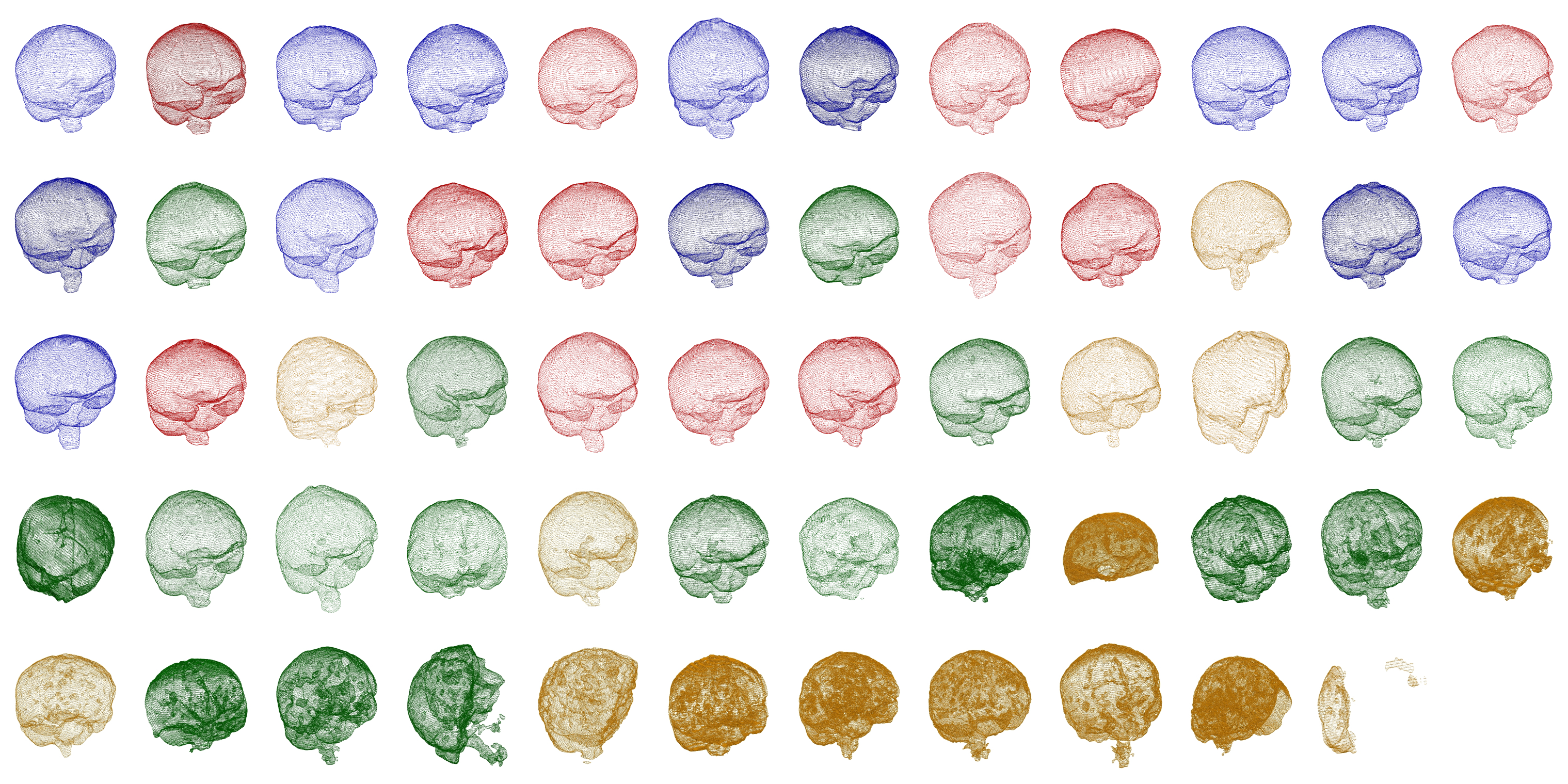}\\[-9pt]
  {\color{black!30}\rule{.7\linewidth}{0.2pt}}\\[1.5pt]
  \makebox[\linewidth][l]{\small (b)}\\[-11pt]
  \includegraphics[width=.7\linewidth]{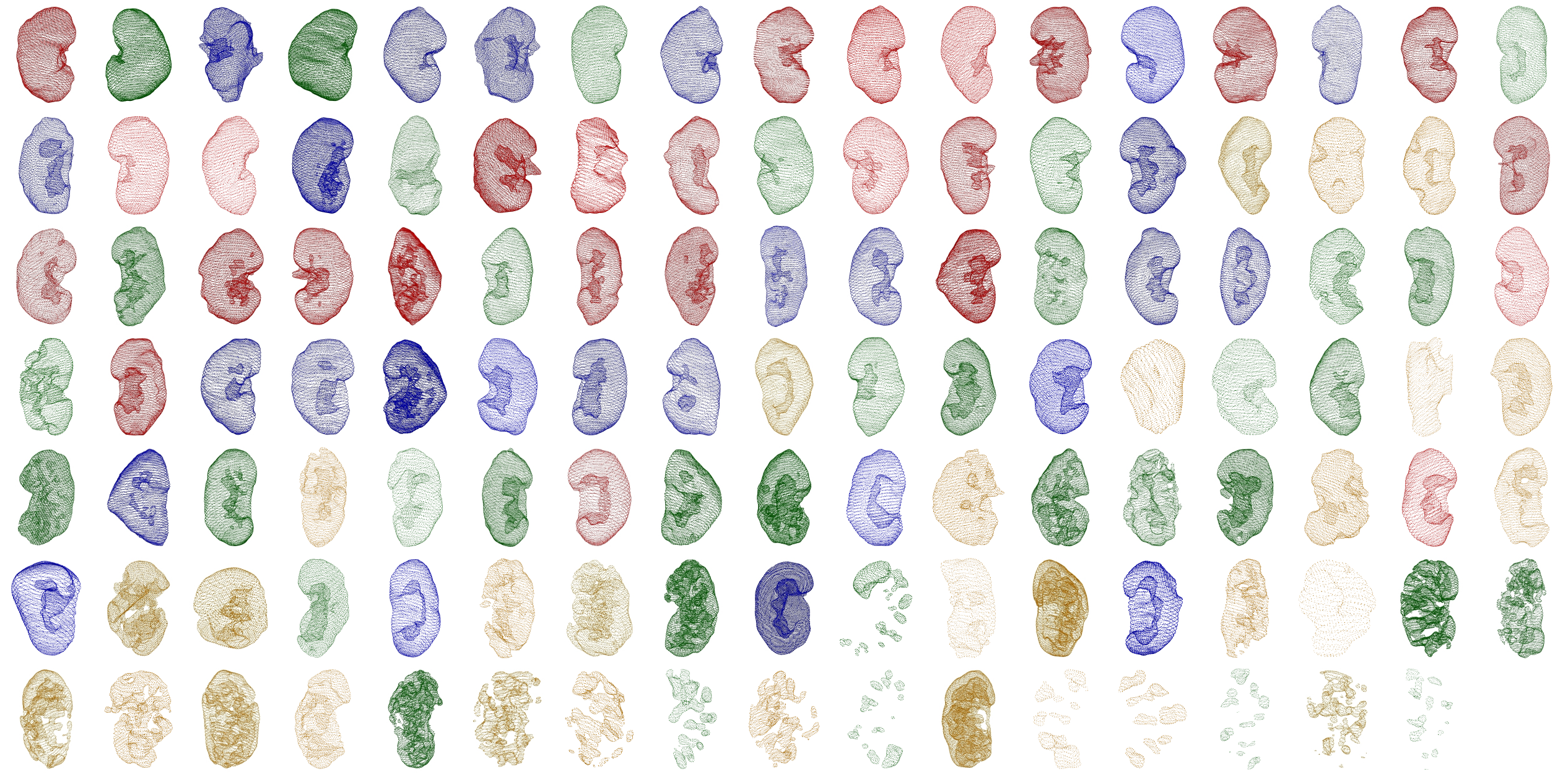}
  \caption{Decomposition atlases for (a) 59 brains, ordered by Borda aggregation of $d^{\mathrm{fus}}$-based seriation and PNC rank scores; and (b) 118 kidneys, by seriation of $d^{\mathrm{fus}}$. Each atlas reads left to right and top to bottom, from intact to severely decomposed; colors show the whole-body stage as in Fig.~\ref{fig:br_3d_orig}.}
  \label{fig:atlas}
\end{figure}

To illustrate how Borda rank aggregation balances between the TDA-based and PNC-based ranks, we present a graphical illustration in Fig.~\ref{fig:rank_br_comb}. 
Each point corresponds to a brain sample, and its orthogonal projection onto the diagonal indicates the Borda-aggregated rank scores. 
Along the diagonal, observations closer to the origin are ranked earlier in the final atlas. Borda ties (e.g., Brains~10 and~26) are broken by the absolute difference between the two rank scores, with samples with larger differences placed later. The resulting decomposition atlas for the brains is shown in Fig.~\ref{fig:atlas}(a).

Most brains lie within one standard deviation around the diagonal, indicating close agreement between the two orderings; only a few fall outside the $\pm 2$~SD band. These discordant cases illustrate the complementary information captured by the two views. For example, Brains~30 and 39 are ranked late by PNC but not by TDA. They have relatively small volumes but comparatively simple internal structures, explaining the discrepancy between the two rankings.


\subsection{Kidney atlas}

Registration did not yield meaningful geometric results for kidneys (Section~\ref{sec:pca_pnc_res}), so the kidney atlas is based on the topological ordering alone.
Figure~\ref{fig:atlas}(b) shows the seriation of $d^{\mathrm{fus}}$. Kidneys early in the ordering generally retain intact structures with well-defined outer shells; then the shapes gradually exhibit increasing structural degradation, including growing outer boundary deformation and the emergence of interior cavities; toward the end of the atlas, the samples become substantially degraded, with only sparse or fragmented remnants.

\subsection{Cross-organ consistency}

\begin{figure}[ht!]
    \centering   
    \includegraphics[width=0.65
         \linewidth]{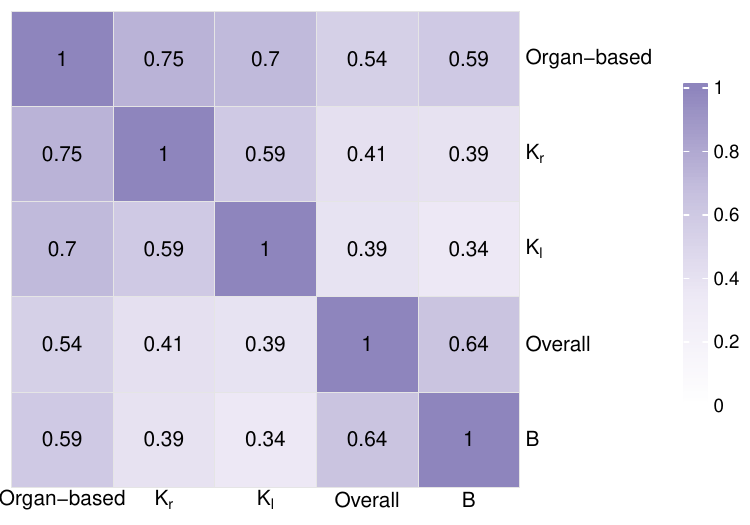}
    \caption{Kendall’s $\tau$ among organ-specific rank scores ($B$ for brains, $K_l$ for left kidneys, and $K_r$ for right kidneys), whole-body decomposition stage (overall), and organ-based decomposition score (organ-based).
    }
    \label{fig:rank_organ}
\end{figure}

Comparing the atlas positions of the brain and kidneys from the same decedent allows us to test whether decomposition is synchronized across organs.
For each decedent, we record the brain rank (\(B\)), the left- and right-kidney ranks (\(K_l\) and \(K_r\)), the Borda aggregate of these three ranks, which summarizes internal organ condition, and the whole-body decomposition stage. Figure~\ref{fig:rank_organ} reports pairwise Kendall's $\tau$ values. Three patterns stand out:
\begin{itemize}
    \item Paired kidneys from the same individual agree only moderately in decomposition stage (Kendall's $\tau = 0.59$), so two organs of the same type within one body can differ in decay status. This difference, however, shows no systematic side bias: left kidneys were neither more nor less decomposed than right kidneys (Fig.~\ref{fig:wilcox} in Section~\ref{sec:LRKidney}).
    \item Agreement across organ types is weaker than that of paired organs ($\tau = 0.34$ between $B$ and $K_l$, $0.39$ between $B$ and $K_r$): decomposition differs more between organ types than within them. 
    \item The whole-body decomposition stage shows moderate agreement with the brain ordering ($\tau = 0.64$) but only weak agreement with the left- and right-kidney orderings ($\tau = 0.39$ and $0.41$, respectively). Its agreement with the aggregated organ-based rank is intermediate ($\tau = 0.54$). These patterns indicate that internal organs decompose asynchronously: whole-body assessment captures some of this variation but largely misses kidney-specific decomposition.
\end{itemize}


\section{Discussion}
\label{sec:discuss}

This study used whole-body CT scans from NMDID and statistical shape analysis of segmented brains and kidneys to characterize internal postmortem decomposition at the organ level. The organ-specific orderings suggest that whole-body staging does not fully capture the variation in internal organ morphology. This variation includes both cavitation and tissue loss, which complicate comparisons based on anatomical correspondence. The atlases provide reference collections for examining these differences.

\subsection{Methodological implications for shape analysis}

Whether registration is meaningful varied by organ in this study. When organs share a recognizable common structure, as brains do, registration-based methods exploit coordinate-level geometry and add information that topology lacks. However, when no single template fits every sample due to the presence of interior cavities, missing tissue, and fragments, as with kidneys, forcing one introduces artifacts such as the over-representation of void points (Section~\ref{sec:pcapnc}). Topological summaries avoid the need for correspondence at the price of some geometric detail, a trade that was decisively favorable for kidneys and roughly neutral for brains. These findings suggest a practical strategy for other shape collections with severe structural heterogeneity, whether arising from degradation, pathology, or development: first assess whether registration is meaningful, and then choose the method family accordingly.

Two further choices that deserve comment and may transfer in other studies: (1) Feature fusion, since no single topological summary captured all relevant variation. Diagram-based distances and distances between functional summaries emphasize different aspects of persistent homology. Standardized fusion of complementary distances and features is a simple construction with no additional tuning once the component distances are chosen, and in simulations it improved both accuracy and stability relative to the best individual distance. (2) Choose the component distances in a mechanistic simulation before analyzing the real data to guard against picking topological summaries by their agreement with visual expectation. The simulation was built to resemble the target degradation process, so the distance and feature combination was selected and fixed before the real-data rankings were constructed.

\subsection{Implications for forensic science}
\label{sec:implication_forensic}

Section~\ref{sec:atlas} gives quantitative form to a concern that the taphonomy literature has mostly stated qualitatively \citep{zhou2011factors}: external assessments are imperfect summaries of internal decompositions. In our cohort, the whole-body score agreed moderately with the brain ordering but only weakly with either kidney ordering, and organs within a body disagreed among themselves. Anatomy offers one possible explanation. The brain decomposes inside the rigid, closed skull, which may buffer external influences and make its decay more reproducible, whereas the kidneys are more exposed to variable routes of autolysis and bacterial access through the vasculature and urinary tract, consistent with the hilum-centered cavities we repeatedly observed. Whatever the mechanism, a single external whole-body score should not be treated as a proxy for the state of any particular internal organ, and organ-level measurement such as the one developed in this study complements whole-body instruments such as the total body score \citep{megyesi2005using} and the radiological alteration index \citep{egger2012development}.
 
The decomposition atlases are designed as reference collections. By ordering from intact to severely decayed, they provide a standard for placing new cases, a material for training and quality control for the technologists and radiologists, and a way to flag bodies whose external whole-body stage and internal organ state disagree markedly, several of which appeared in our own sample (Section~\ref{sec:morphvar}). Because the pipeline needs no template, placing a new case only requires computing its persistence diagram and its distances to the reference organs.

Postmortem interval estimation is the application that may be of interest to forensic readers. However, as discussed in Section~\ref{sec:cohort}, the recorded delay to scan runs from the declaration of death rather than the biological moment of death in this cohort. For the most decomposed bodies, most of the true postmortem interval precedes the recorded clock. The cohort thus contains no usable measure of time since death. Many intrinsic and extrinsic factors are also known to further complicate this estimate of time since death \citep{zhou2011factors}. Decomposition rate is strongly modulated by temperature and environment \citep{megyesi2005using,galloway1989decay}, and such information is unavailable for our cohort. In addition, each specimen has one postmortem interval but, as we have shown, organs can be in substantially different states, so no single-organ score should be expected to track elapsed time cleanly. Joint modeling of multiple organ trajectories with environmental covariates, made possible by organ-level measurements, is a more promising route.

\subsection{Limitations and future directions}
The study has several limitations. The cohort is relatively small (59 natural-death decedents) and comes from one state with a hot, arid climate \citep{galloway1989decay}, and excluding chemotherapy, drug, and alcohol exposure narrows it further. Patterns may differ under other climates, causes of death, or postmortem treatment. The whole-body stage labels are operational judgments rather than a validated instrument. Because no external gold standard exists for evaluating the atlases, current validation is internal, based on consistency across methods, cluster interpretability, and agreement of patterns across organs. Benchmarking in an external cohort would provide stronger evidence. We used only $H_2$, although connected components ($H_0$) and loops ($H_1$) may carry signal in other studies. Some kidney cavities may reflect pre-existing cysts, particularly in older individuals \citep{rule2012characteristics}, complicating their attribution to postmortem decomposition. Finally, the mechanistic simulation only covers one decay mechanism of progressive cavity formation; a simulation spanning more potential decay modes would provide stronger support for distance choices.

This work also opens several directions. The current bottleneck is the time required to process each case; automated segmentation of decomposed organs would allow the analysis to scale to the thousands of NMDID scans. Extending to additional organs such as the liver, heart, and lungs would also allow us to turn organ-specific atlases into organ-\textit{system} decomposition atlases. Methodological next steps include considering richer invariants such as multiparameter persistence and developing uncertainty measures for seriation-based orderings. Linking organ-level differences to temperature records, circumstances of death, and other environmental history could help incorporate internal morphology into future postmortem interval models, thereby complementing established external instruments \citep{megyesi2005using,egger2012development}. Validation of the atlases as reference material in forensic practice by comparing inter-rater studies with and without atlas support would test their value in practice.

In sum, decomposed organs break the default assumption of shape analysis that observations can be registered to a common template. Persistent homology offers a registration-free alternative that makes internal organ decomposition measurable from postmortem CT. We release the atlases, point clouds, and code so that the considerable cost of constructing these data is paid only once.

\begin{acks}[Acknowledgments]
The authors gratefully acknowledge the computational resources provided by the Hyperion high performance computing cluster and Research Computing at the University of South Carolina (RRID:SCR\_027488).

\end{acks}

\bibliographystyle{imsart-nameyear} 
\bibliography{Bibliography}       

\newpage
\renewcommand{\footnoterule}{}

\begin{frontmatter}
\title{Supplementary Material for Decomposition Atlases of Brains and Kidneys from Postmortem CT: Shape and Topological Data Analysis}
\runtitle{Supplementary Material}

\end{frontmatter}

\setcounter{section}{0}
\setcounter{subsection}{0}
\setcounter{figure}{0}
\setcounter{table}{0}
\setcounter{equation}{0}

\renewcommand{\thefigure}{S\arabic{figure}}
\renewcommand{\thetable}{S\arabic{table}}
\renewcommand{\thesection}{S\arabic{section}}
\renewcommand{\theHfigure}{S\arabic{figure}}
\renewcommand{\theHtable}{S\arabic{table}}
\renewcommand{\theHsection}{S\arabic{section}}

\section{Cohort metadata variables and point cloud sizes}
\label{sec:Meta}
\begin{figure}[ht!]
    \centering
    \includegraphics[width=1\linewidth]{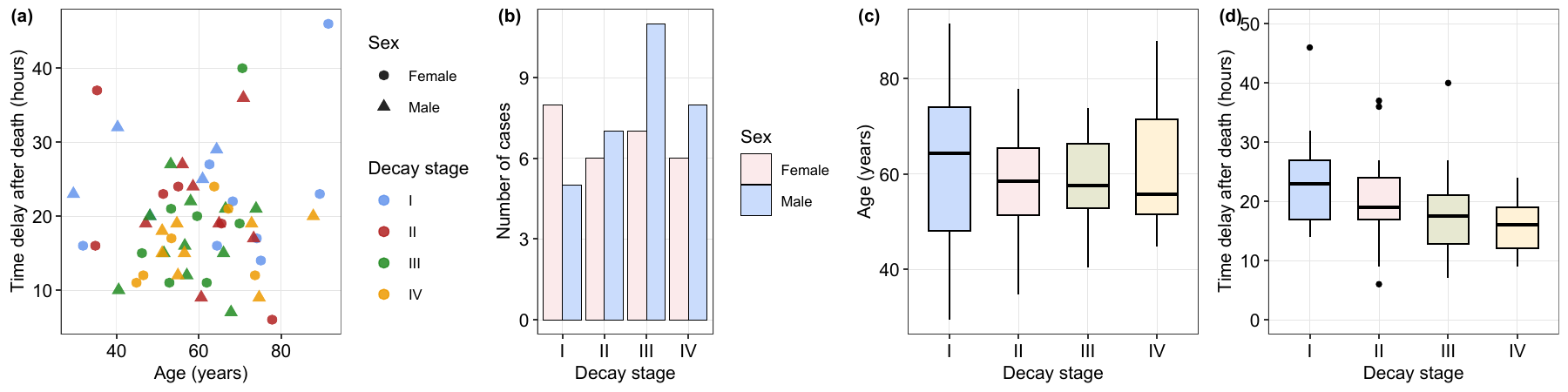}
    \caption{Distribution of decedent sex, age at death, and time delay after death by decay stage.
    (a) Scatter plot of age at death in years and time delay after death to CT scan in hours, colored by decay stage and shaped by sex.
    (b) Bar plot showing the distribution of sex across whole-body stages I--IV.
    (c) Boxplots showing age distribution across whole-body stages I--IV.
    (d) Boxplots showing time delay after death to CT scan across whole-body stages I--IV.
    }
    \label{fig:age_delay_stage}
\end{figure}

Figure~\ref{fig:age_delay_stage} shows the distribution of sex, age, and the time delay after death across whole-body stages. The associated tests are reported in Section~\manualref{2.1} of the main text.

\begin{figure}[ht!]
    \centering
    \includegraphics[width=0.9\linewidth]{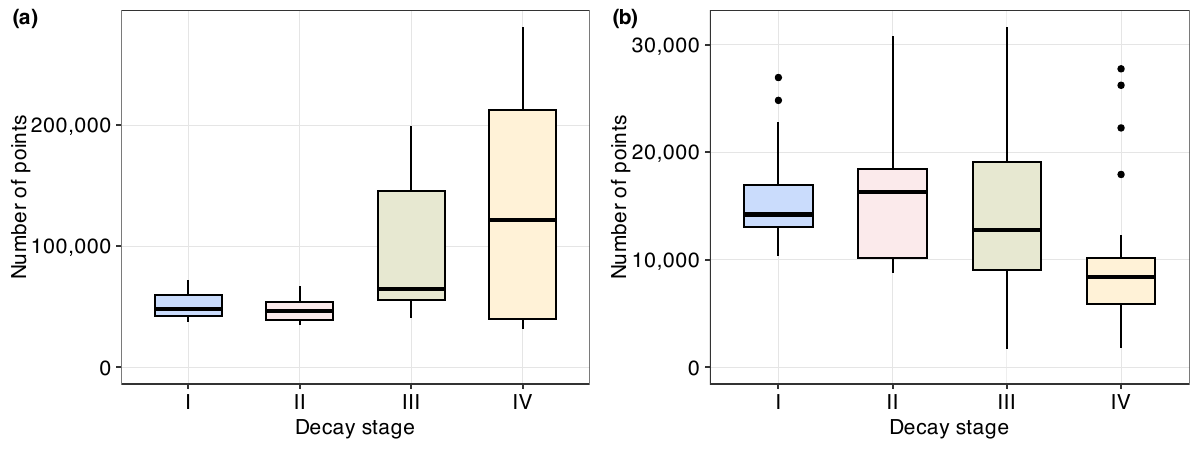}
    \caption{Distribution of point cloud size across (a) brains and (b) kidneys. Roman numerals I--IV indicate the whole-body decomposition stages.}
    \label{fig:ss}
\end{figure}

Figure~\ref{fig:ss} shows the distribution of point cloud sizes for the 59 brains and 118 kidneys, described in Section~\manualref{2.2} of the main text. Brains generally contain more points than kidneys, reflecting differences in organ volume and structural complexity. The number of points in structurally intact organs is represented mainly by their outer surface and therefore has relatively fewer points, as for most Stage~I--II brains in Fig.~\ref{fig:ss}(a). When decay generates intricate internal cavities, the organ is represented by more points, as in some Stage III--IV brains and kidneys. As shown in Fig.~\ref{fig:ss}(b), very small counts can also occur for Stage III--IV organs when so little tissue remains that only fragments are reconstructed.

\section{Registration-based analysis: templates, shape components, and template sensitivity}
\label{sec:Template}
\begin{figure}[ht!]
    \centering    
    \includegraphics[width=0.17\linewidth]{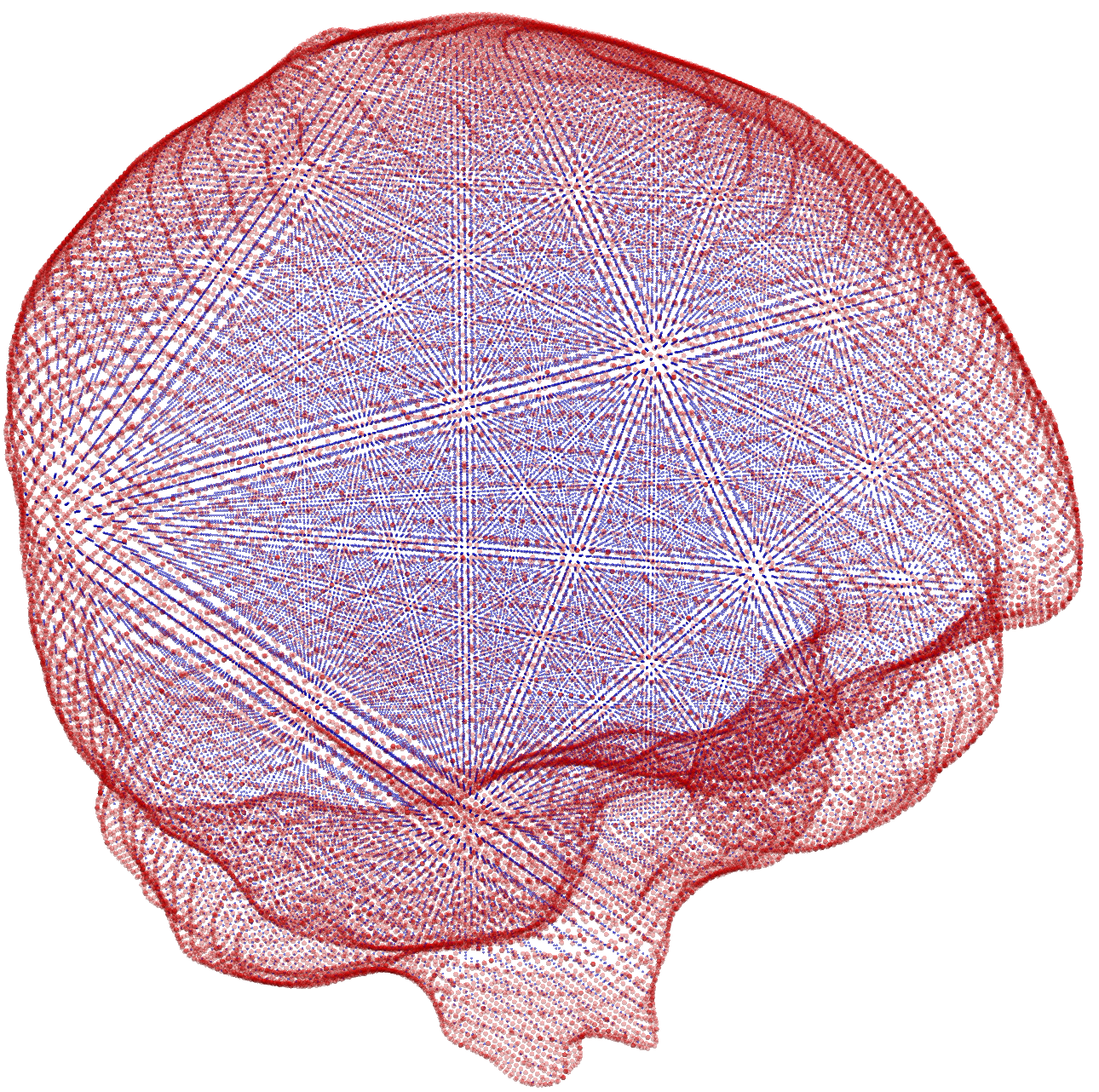}
    \hspace{2.5cm}
    \includegraphics[width=0.1\linewidth]{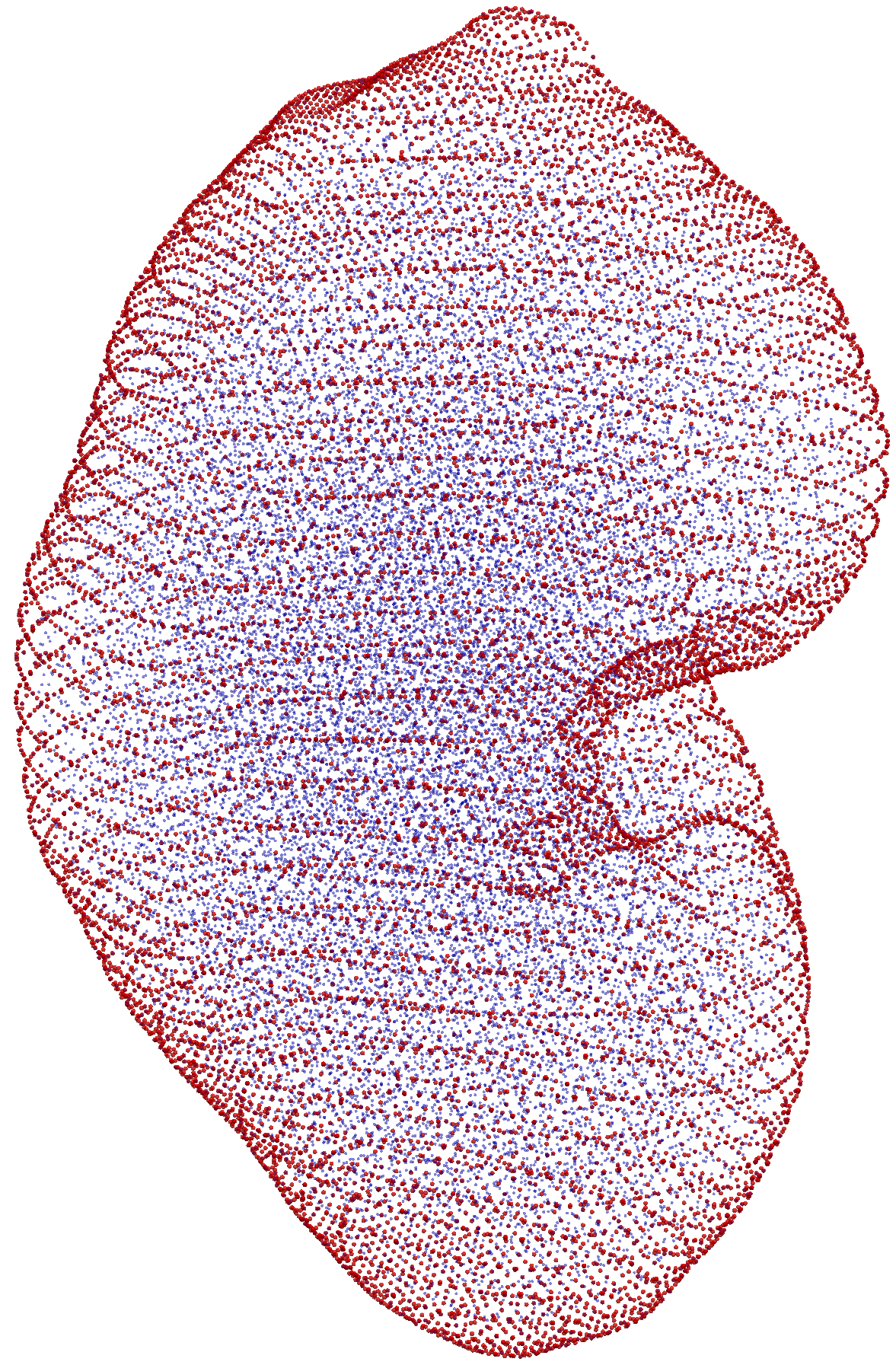}
    \caption{Brain (left) and kidney (right) templates used for ICP registration; surface points red, interior points blue.
    }
    \label{fig:ICPtemp}
\end{figure}

Registration-based methods require a template to which all samples are aligned. Figure~\ref{fig:ICPtemp} shows the brain and kidney templates used to get results in Section~\manualref{3} of the main text. The brain template was constructed from the outer shell of Brain~1 plus interior points on a regular grid with 3~mm spacing in each dimension. The resulting template contains 46{,}006 surface and 53{,}806 interior points. The kidney template is the outer shell of Kidney~1 (13{,}798 points), scaled by a factor of 1.3 to accommodate kidneys that expand during early decomposition. Its 14{,}202 interior points were generated by shrinking the outer shell inward and thinning the result through farthest point sampling \citep{eldar1997} to achieve an approximately uniform distribution of interior points. 

\begin{figure}[ht!]
    \centering    
    \includegraphics[width=.9\linewidth]{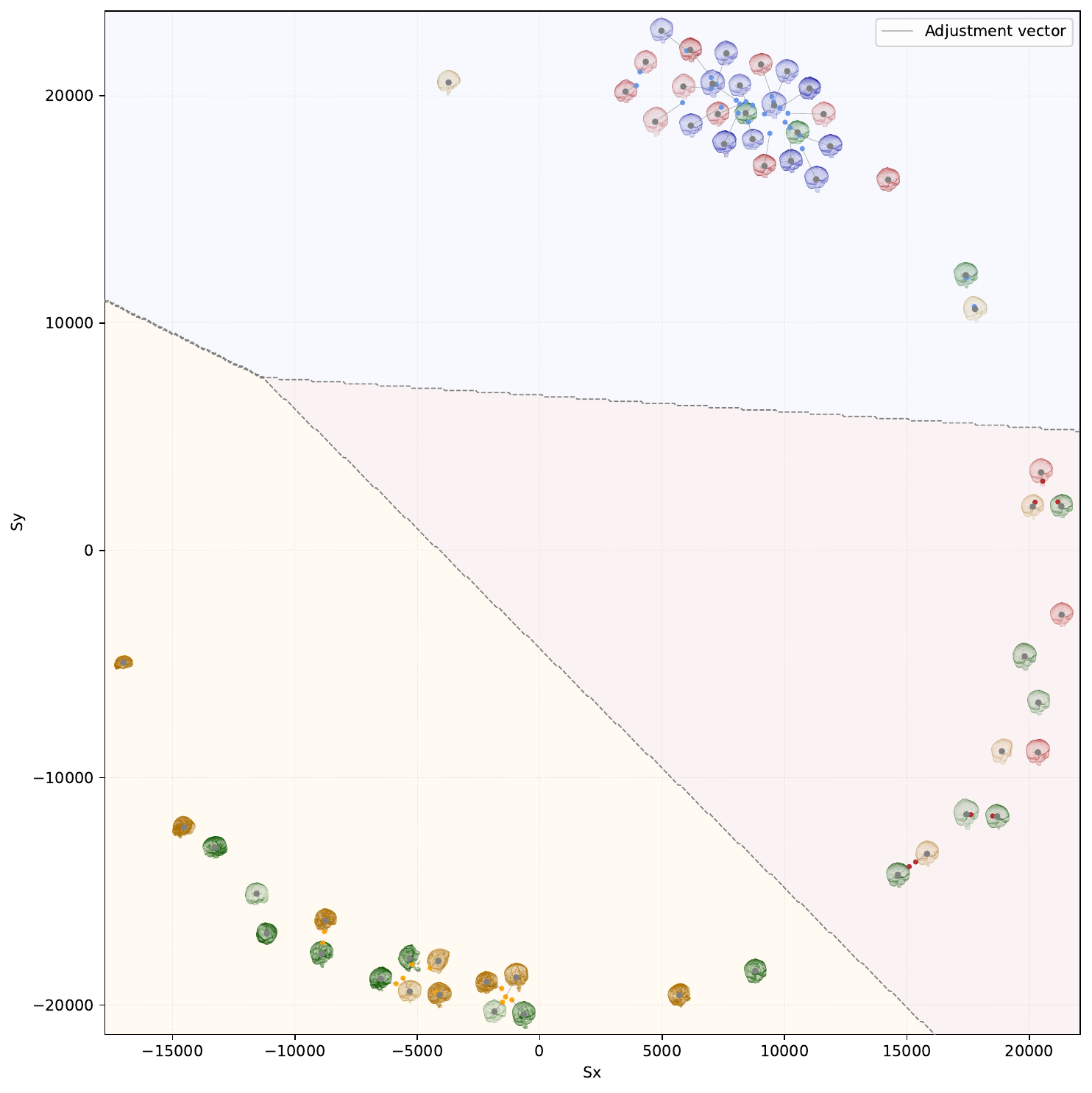}
    \caption{Brains embedded by fast PNC polar scores, colored by whole-body stage labels.
    }
    \label{fig:br_pnc_polar}
\end{figure}

\begin{figure}[ht!]
    \centering    
    \includegraphics[width=0.8\linewidth]{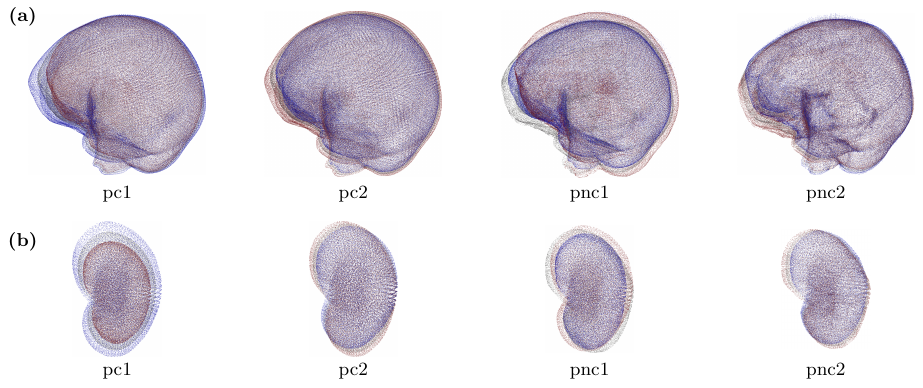}
    \caption{Shape variation in (a) the brain and (b) the kidney along the first two PCA and fast PNC components. For each component, shapes are reconstructed at the mean score minus one standard deviation (blue), the mean score (gray), and the mean score plus one standard deviation (red).
    }
    \label{fig:bf_br_kd}
\end{figure}

Figure~\ref{fig:br_pnc_polar} shows the brain embeddings obtained using fast PNC polar scores. The embeddings are ordered by decomposition severity along an arc spanning approximately three-quarters of a circle. 
For each of the first two components, shapes were backfitted at the mean and at one standard deviation on either side (mean $\pm$ 1\,SD), with the results shown in Fig.~\ref{fig:bf_br_kd}. For both brains and kidneys, PC1 primarily captures size variation, whereas PNC1 captures variation in both size and shape. This pattern is consistent with the higher percentage of variation explained by PNC1 than by PC1 in both organs.


\begin{figure}[ht!]
    \centering   
    \includegraphics[width=1\linewidth]{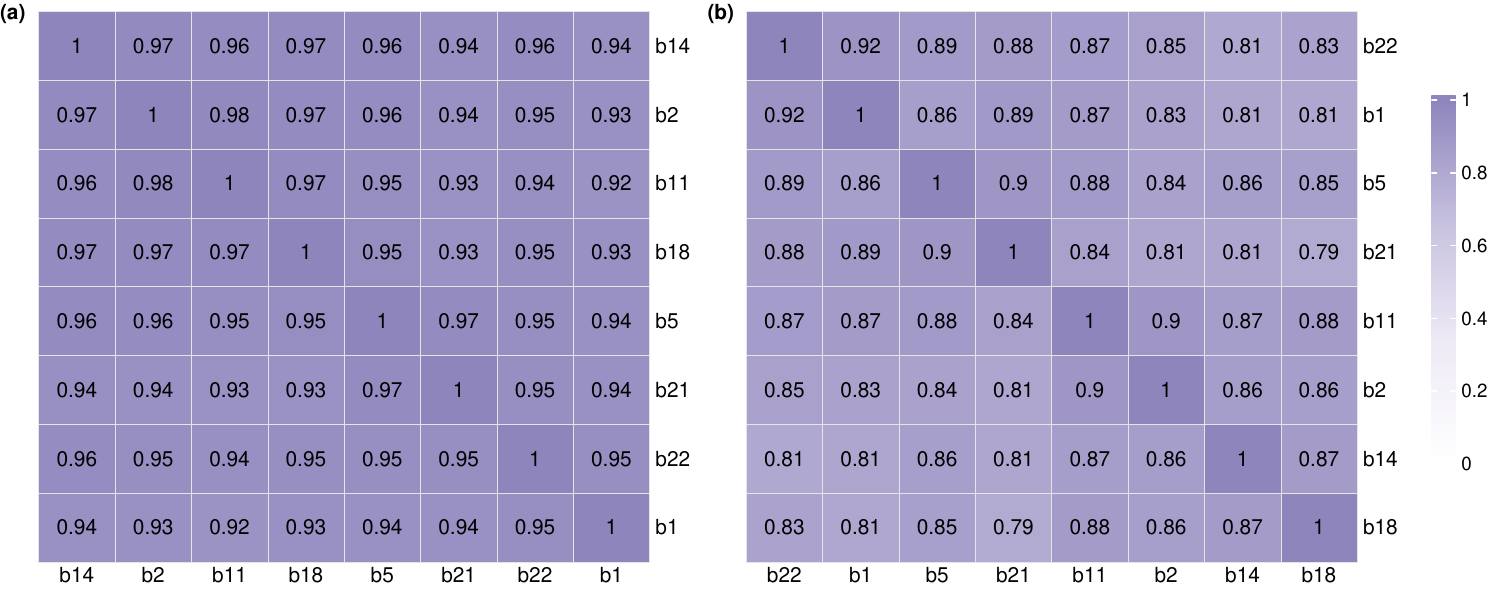}
    \caption{Rank agreement across template choices for brain samples using (a) PCA and (b) PNC. Rows and columns correspond to different templates. Heatmap values show Kendall's $\tau$.
    }
    \label{fig:br_templates}
\end{figure}

\begin{figure}[ht!]
    \centering   
    \includegraphics[width=1\linewidth]{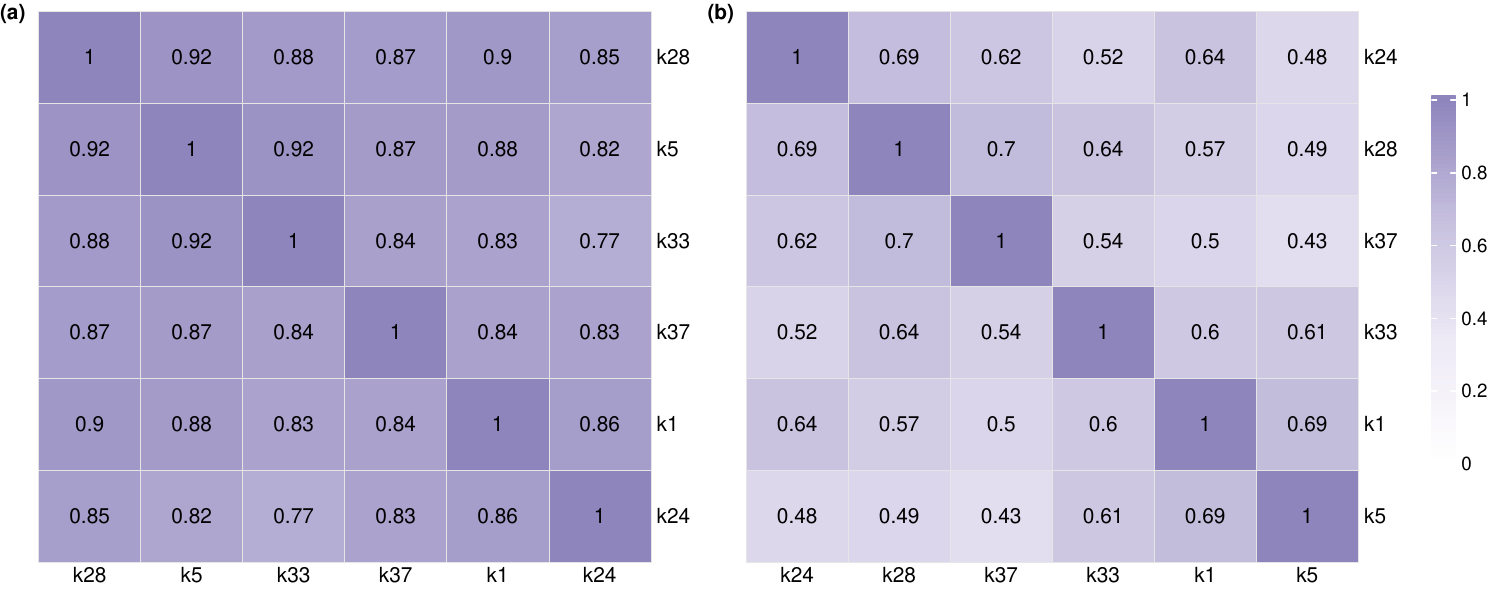}
    \caption{Rank agreement across template choices for kidney samples using (a) PCA and (b) PNC. Rows and columns correspond to different templates. Heatmap values show Kendall's $\tau$.
    }
    \label{fig:kd_templates}
\end{figure}

To assess sensitivity to template choice, we repeated the registration and ranking with alternative templates. For brains, eight templates were constructed from Brains~1, 2, 5, 11, 14, 18, 21, and 22; for kidneys, templates were constructed from Kidneys~1, 5, 24, 28, 33, and 37. For each template, the outer and inner template nodes were generated as described above. All samples were independently registered to each template by rigid ICP and nearest-neighbor matching. PCA and PNC were then applied to the registered data, and samples were ranked according to their PC1 and PNC1 scores. Rank agreement between each pair of templates was measured using Kendall’s \(\tau\). The results are shown in Figs.~\ref{fig:br_templates} and \ref{fig:kd_templates}. For brains, all pairwise values are at least 0.79, which indicates that the geometric ordering is not driven by the particular template. For kidneys, agreement is substantially lower, which indicates that the registration-based ordering depends on the template. This instability is another reason the kidney atlas in the main text is constructed from the registration-free analysis alone.

\section{Choice of homology dimension}
\label{sec:Dimension}

\begin{figure}[ht!]
    \centering   
    \includegraphics[width=1\linewidth]{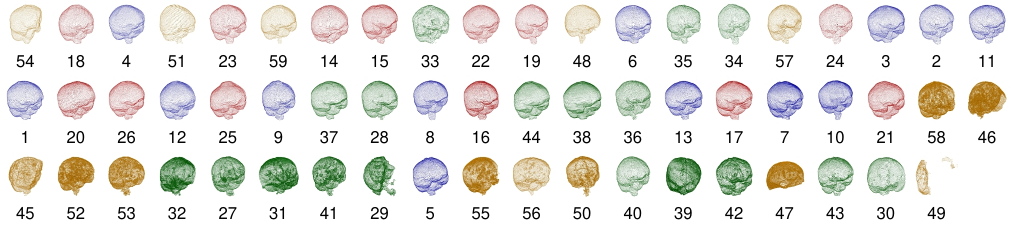}
    \caption{Brain samples ordered by seriation of the fused distance computed from $H_0$. All six individual distances listed in Table~1 of the main text were included in the fusion.
    }
    \label{fig:h0_brain}
\end{figure}

\begin{figure}[ht!]
    \centering   
    \includegraphics[width=1\linewidth]{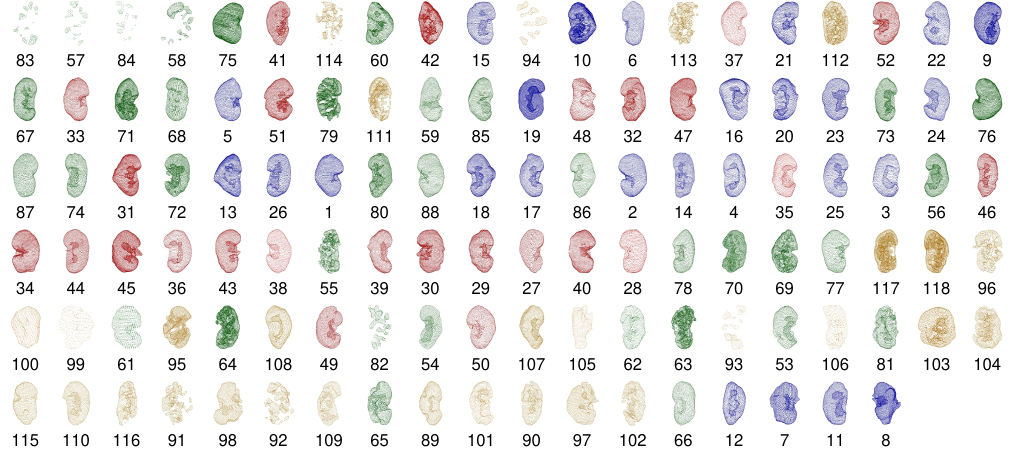}
    \caption{Kidney samples ordered by seriation of the fused distance computed from $H_0$. All six individual distances listed in Table~1 of the main text were included in the fusion.
    }
    \label{fig:h0_kidney}
\end{figure}

\begin{figure}[ht!]
    \centering   
    \includegraphics[width=1\linewidth]{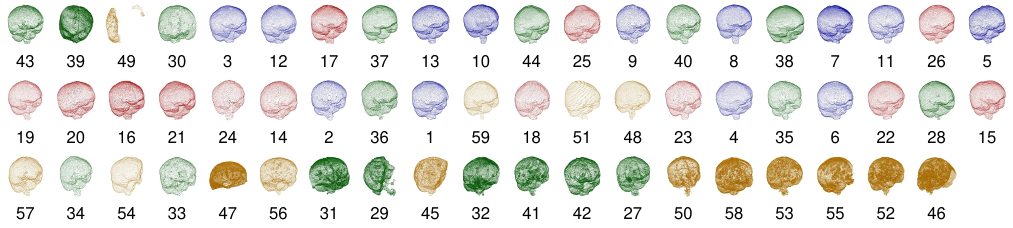}
    \caption{Brain samples ordered by seriation of the fused distance computed from $H_1$. All six individual distances listed in Table~1 of the main text were included in the fusion.
    }
    \label{fig:h1_brain}
\end{figure}

\begin{figure}[ht!]
    \centering   
    \includegraphics[width=1\linewidth]{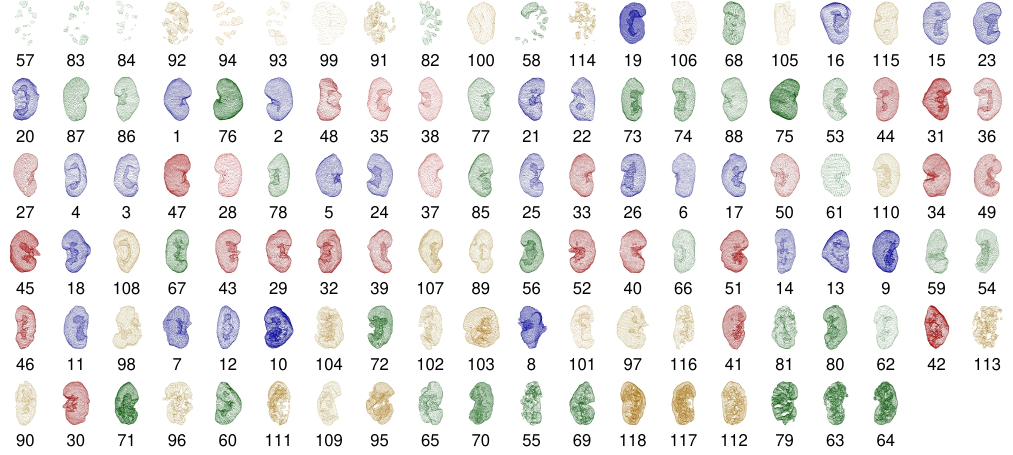}
    \caption{Kidney samples ordered by seriation of the fused distance computed from $H_1$. All six individual distances listed in Table~1 of the main text were included in the fusion.
    }
    \label{fig:h1_kidney}
\end{figure}

\begin{figure}[ht!]
    \centering   
    \includegraphics[width=1\linewidth]{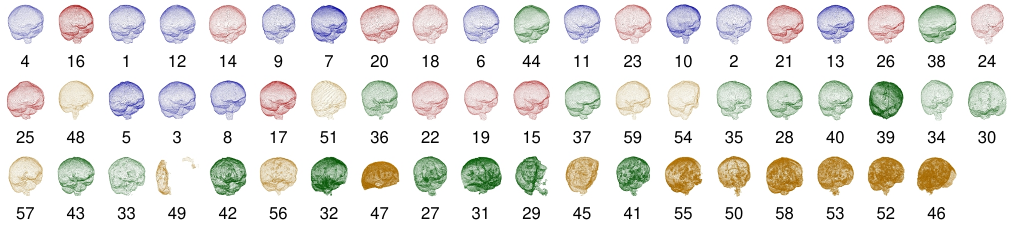}
    \caption{Brain samples ordered by seriation of the fused distance computed from $H_2$. All six individual distances listed in Table~1 of the main text were included in the fusion.
    }
    \label{fig:h2_brain}
\end{figure}

\begin{figure}[ht!]
    \centering   
    \includegraphics[width=1\linewidth]{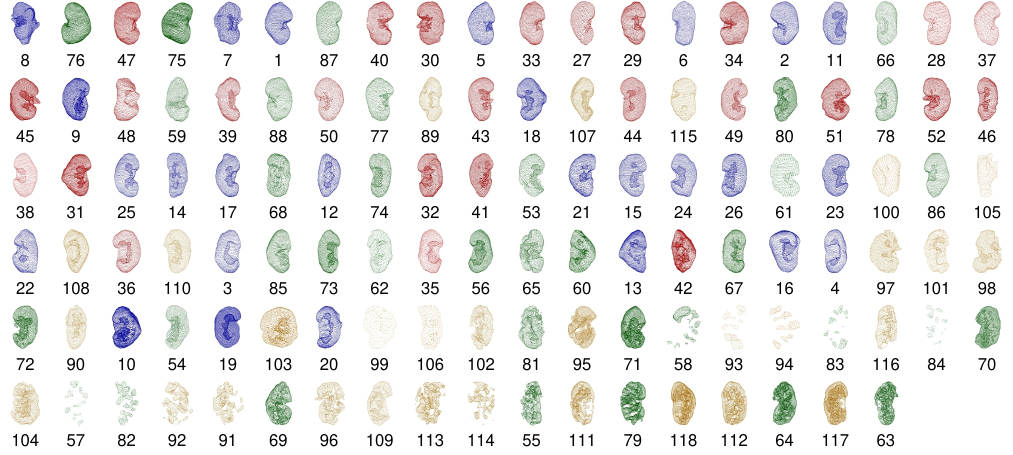}
    \caption{Kidney samples ordered by seriation of the fused distance computed from $H_2$. All six individual distances listed in Table~1 of the main text were included in the fusion.
    }
    \label{fig:h2_kidney}
\end{figure}

Visual inspection of all samples showed that interior cavities were among the most prominent features of the decomposed organs, motivating the use of the second homology dimension, \(H_2\). We also evaluated the performance of \(H_0\) and \(H_1\) before focusing on \(H_2\). Brain and kidney samples were ordered using fused distance matrices derived separately from the three homology dimensions. The results are shown in Figs.~\ref{fig:h0_brain}--\ref{fig:h2_kidney}.

The $H_0$ and $H_1$ orderings do not follow the degree of decomposition. In Figs.~\ref{fig:h0_kidney} and~\ref{fig:h1_kidney}, several heavily fragmented kidneys (e.g., Kidneys~83, 57, 84, and 58) appear near the relatively intact end of the ordering, whereas some visibly intact Stage~I and II kidneys (e.g., Kidneys~50, 7, 30, and 11) appear in the middle or near the advanced-decomposition end. In Fig.~\ref{fig:h1_brain}, the most severely decayed brain, Brain~49, is placed third on the intact end of the ordering. These patterns are consistent with the lifetimes of $H_0$ and $H_1$ features being driven mainly by sampling density in the boundary-sampled point clouds, which reflects more the reconstruction and processing differences rather than decomposition. In contrast, the $H_2$ orderings in Figs.~\ref{fig:h2_brain} and~\ref{fig:h2_kidney} progress from intact to severely decomposed samples. The forensic expert of the team reviewed the three sets of orderings and considered only the $H_2$ orderings anatomically meaningful in the forensic science domain.

\section{Denoising threshold selection}
\label{sec:Denoising}

\begin{figure}[ht!]
    \centering   
    \includegraphics[width=.9\linewidth]{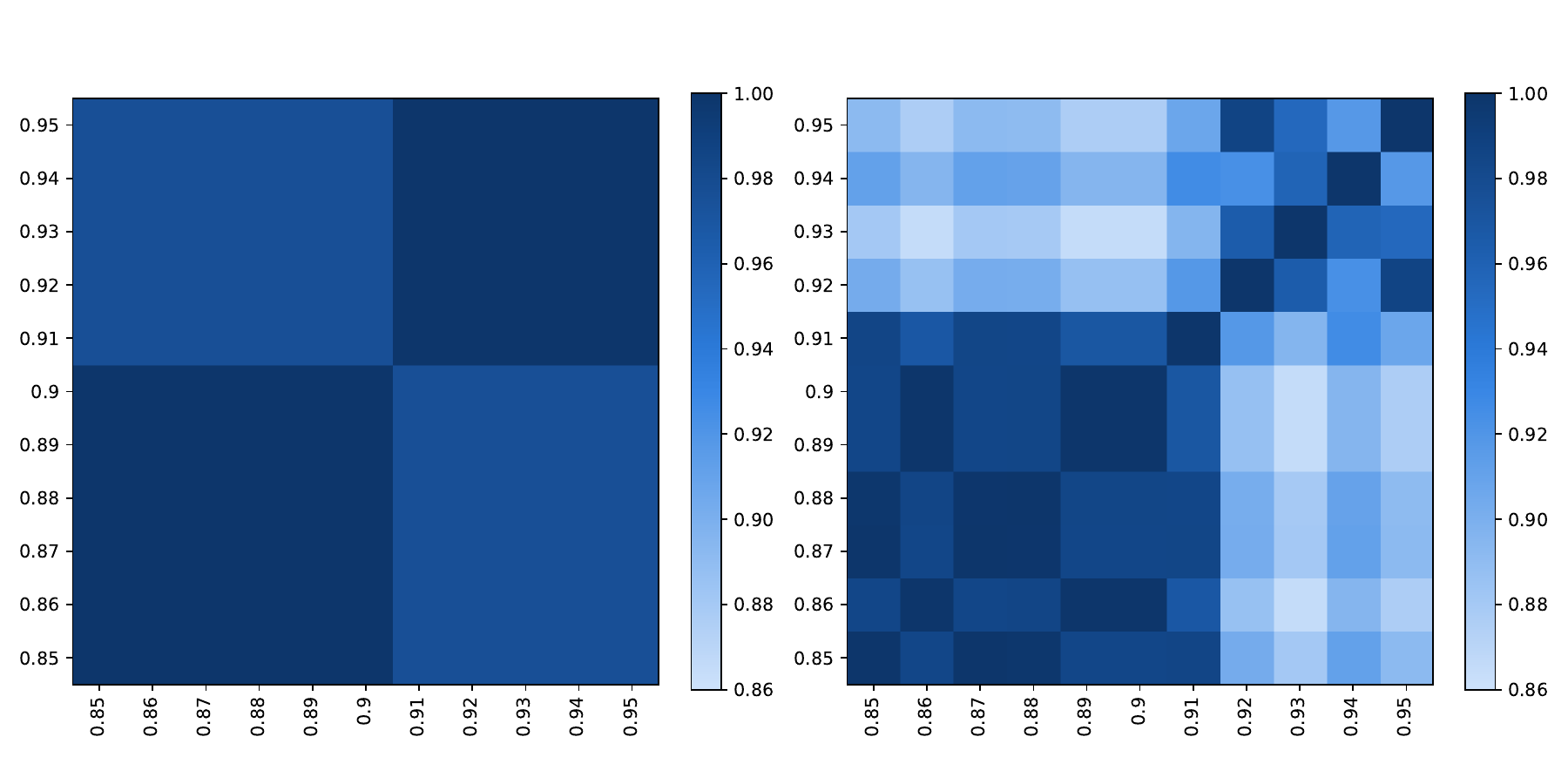}
    \caption{Rank agreement across persistence thresholds for brain (left) and kidney (right) samples during persistence diagram denoising. Rows and columns correspond to percentiles of persistence pooled for each organ. Heatmap values show Kendall's $\tau$. Seriation was applied to $d^\mathrm{fus}$ derived from $d^{\mathrm{B}}$ and $d^{\phi}$.
    }
    \label{fig:thr}
\end{figure}

For each organ, a global persistence threshold was applied to all samples during persistence diagram denoising. The threshold was defined as a percentile of the pooled persistence values from all samples of that organ. To assess sensitivity to the percentile choice, we varied the percentile from 0.85 to 0.95 in increments of 0.01. Sample ranks were recalculated at each percentile, and rank agreement across percentile choices was measured using Kendall’s \(\tau\). Results for the brain and kidney datasets are shown in Fig.~\ref{fig:thr}. Brain orderings agree closely at every threshold; kidney orderings agree closely within the 85th to 91st and within the 92nd to 95th percentiles. The chosen 90th percentile lies inside the stable range.

\section{Stability under coordinate noise}
\label{sec:Stability}

\begin{figure}[ht!]
    \centering   
    \includegraphics[width=0.65\linewidth]{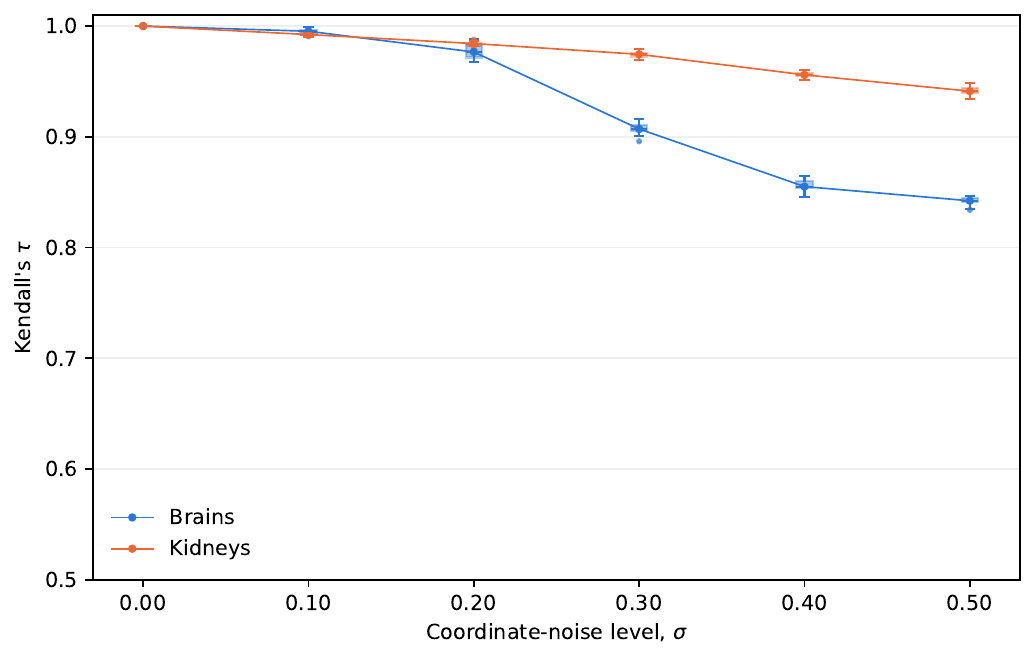}
    \caption{Rank agreement between the original data and data with added Gaussian noise for brain (blue) and kidney (orange) samples. 
    Rank scores from the original data served as the reference and agreement with the noise-added data was quantified using Kendall's $\tau$. Seriation was applied to $d^\mathrm{fus}$ derived from $d^{\mathrm{B}}$ and $d^{\phi}$.
    }
    \label{fig:noise}
\end{figure}

To assess the effect of sampling noise in the reconstructed 3D coordinates, we added Gaussian noise sampled from \(\mathcal{N}(0,\sigma^2)\) to the original data and applied the same analysis described in the main text to the resulting noisy data. For both organs, we considered \(\sigma \in \{0.1, 0.2, 0.3, 0.4, 0.5\}\) mm and repeated the analysis 20 times at each noise level. For each level, we calculated Kendall's \(\tau\) between the sample ranks obtained from the original and noisy data. The results are shown in Fig.~\ref{fig:noise}. Agreement gradually decreases as \(\sigma\) increases for both organs.

\section{Additional figures for the topological analysis}
\label{sec:AdditionalTDA}

\begin{figure}[ht!]
    \centering
    \includegraphics[width=.85\linewidth]{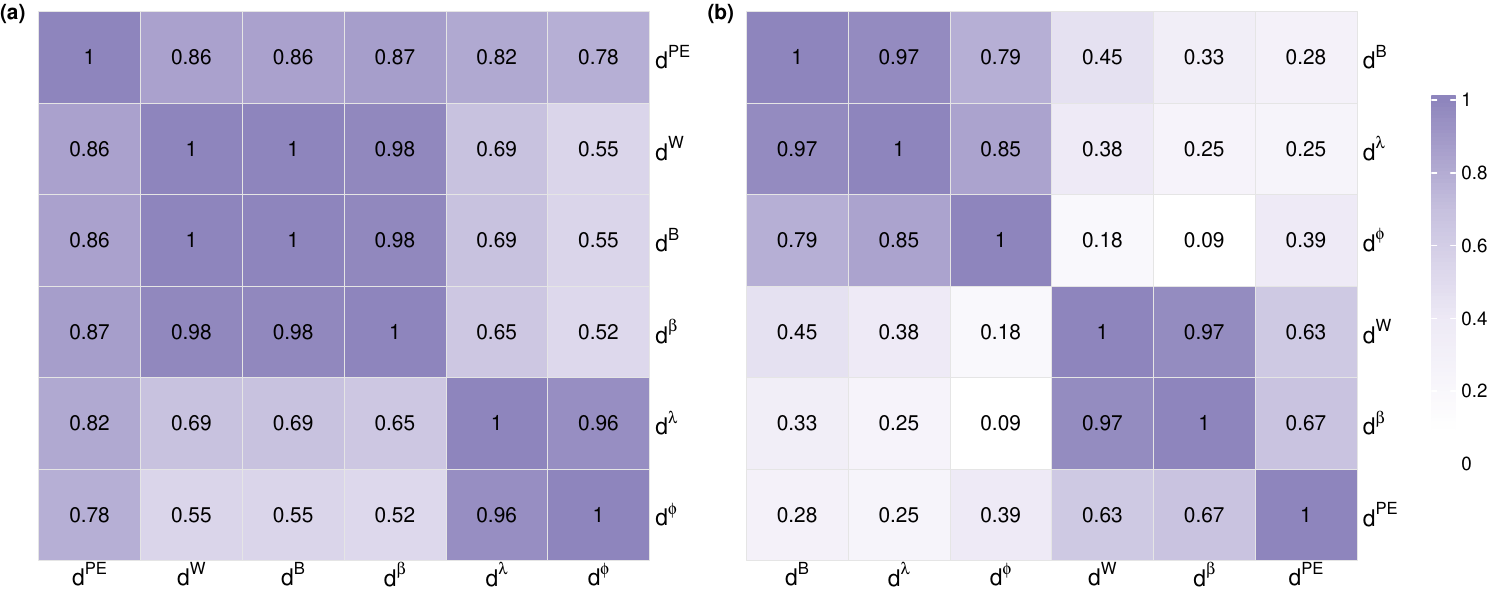}
    \caption{Mantel statistics between different distance matrices for (a) brains and (b) kidneys. 
    }
    \label{fig:compare}
\end{figure}

Figure~\ref{fig:compare} shows the Mantel statistics between the six topological distance matrices, discussed in Section~\manualref{4.3} of the main text. 

\begin{figure}[ht!]
    \centering
    \includegraphics[width=\linewidth]{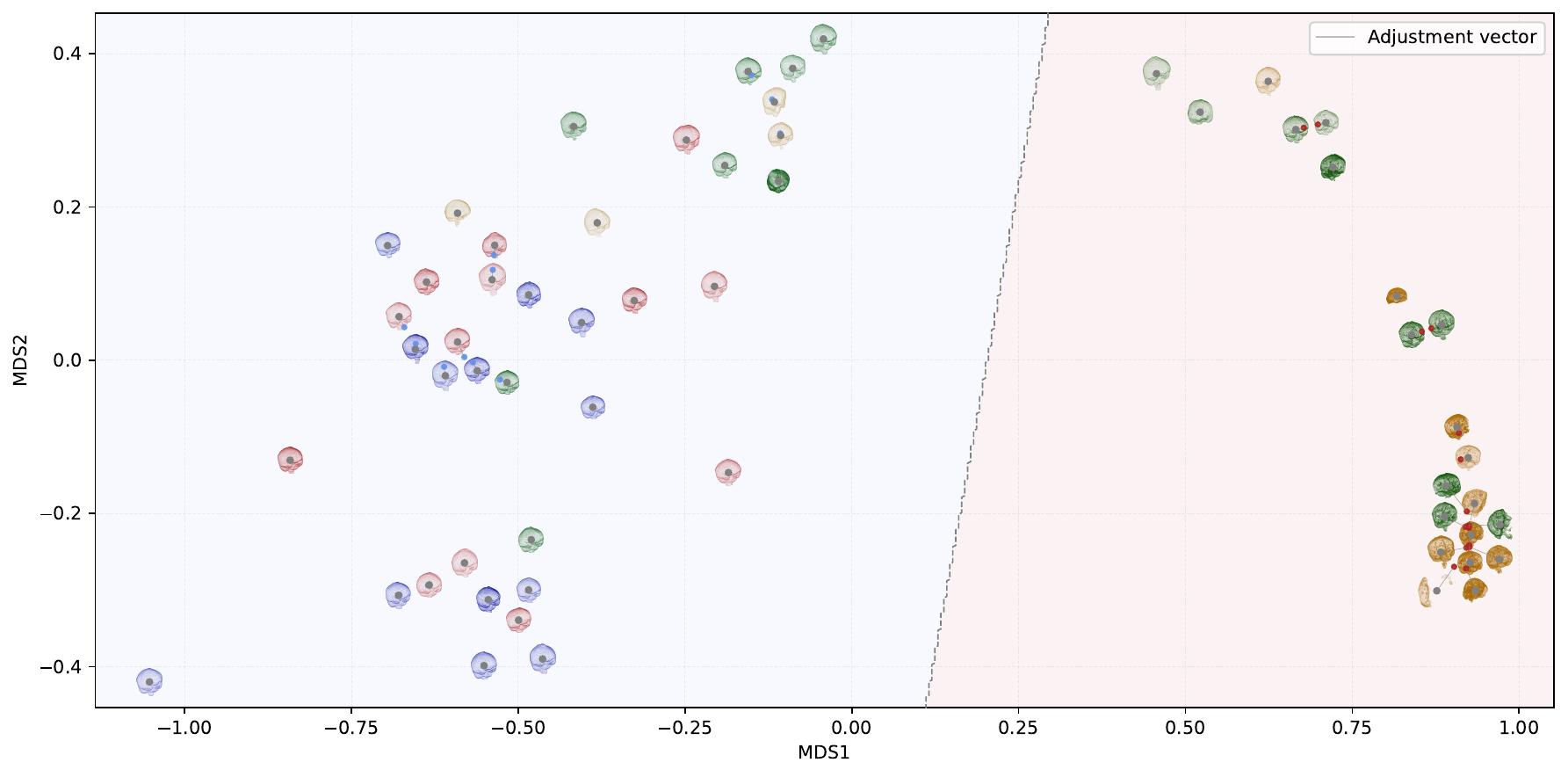}
    \includegraphics[width=\linewidth]{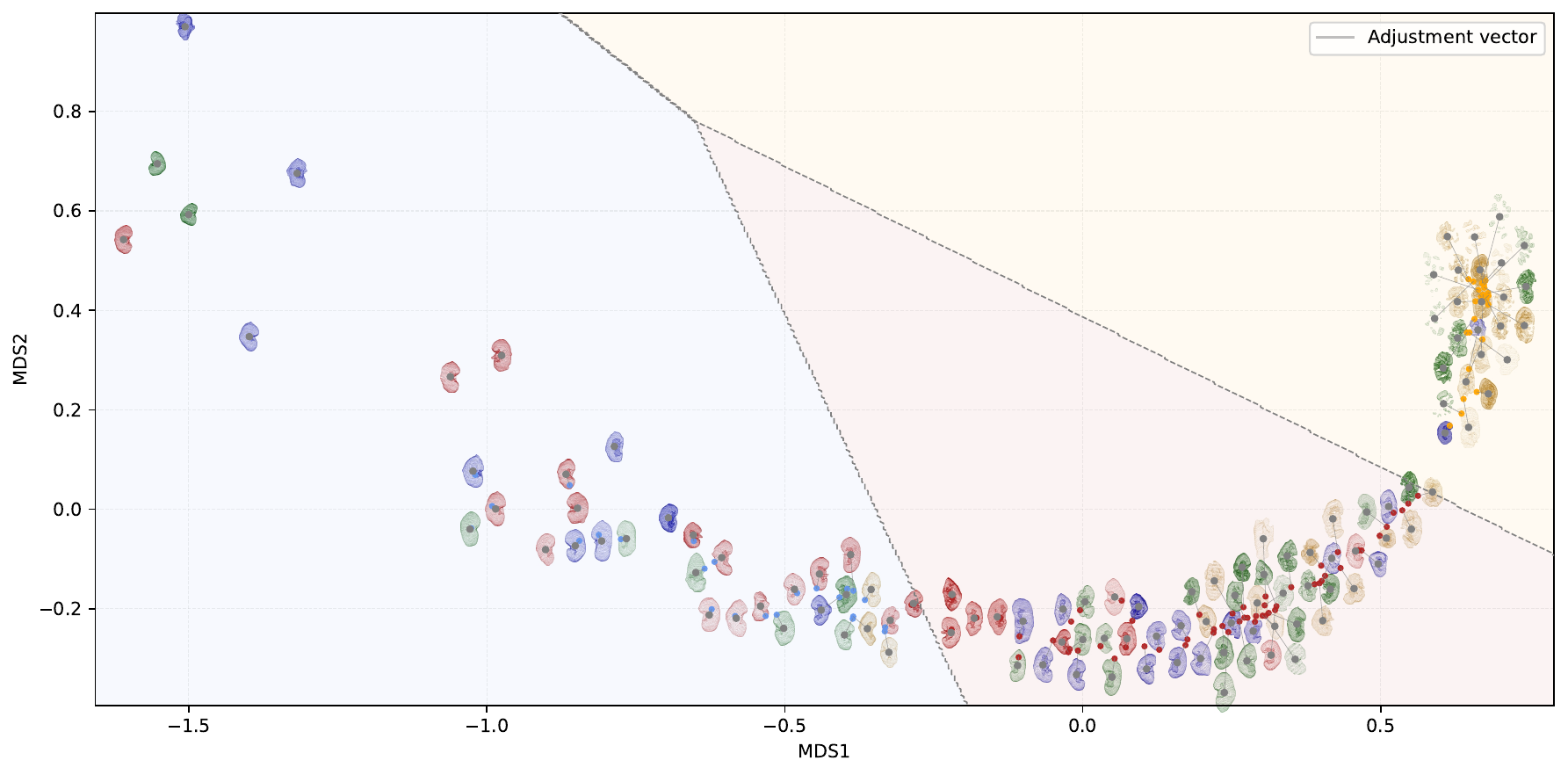}
    \caption{MDS embedding of the fused distance $d^{\mathrm{fus}}$ derived from $d^{\mathrm{B}}$ and $d^{\phi}$ for brains (top) and kidneys (bottom). Each organ is drawn at its embedding position and colored by whole-body stage. Positions are shifted slightly where samples overlap, with gray markers at the original locations. Shaded regions indicate $k$-medoids clusters, with boundaries drawn by LDA for visualization only.}
    \label{fig:tda_br_kd_mds}
\end{figure}

\begin{figure}[ht!]
    \centering    
    \includegraphics[height=4cm]{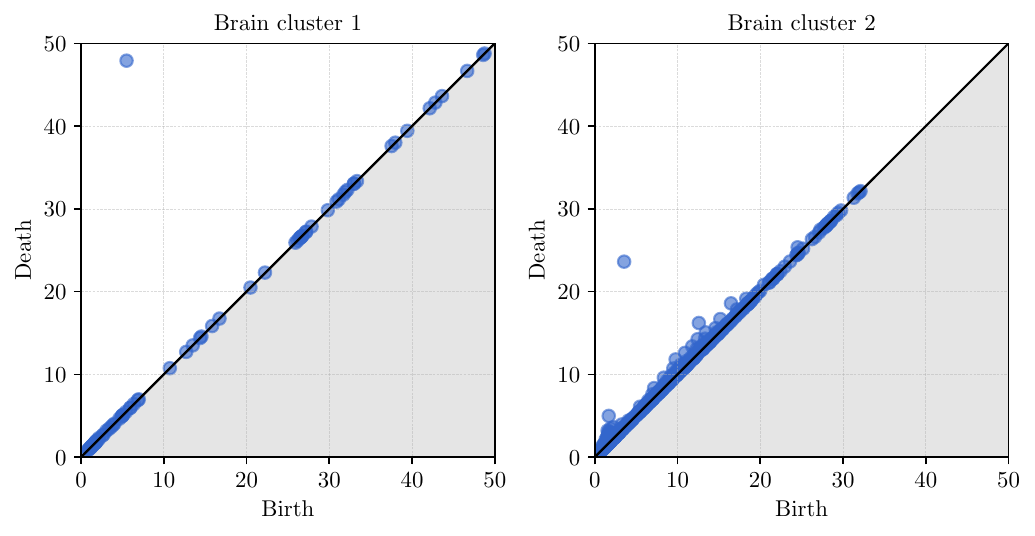}
    \includegraphics[height=4cm]{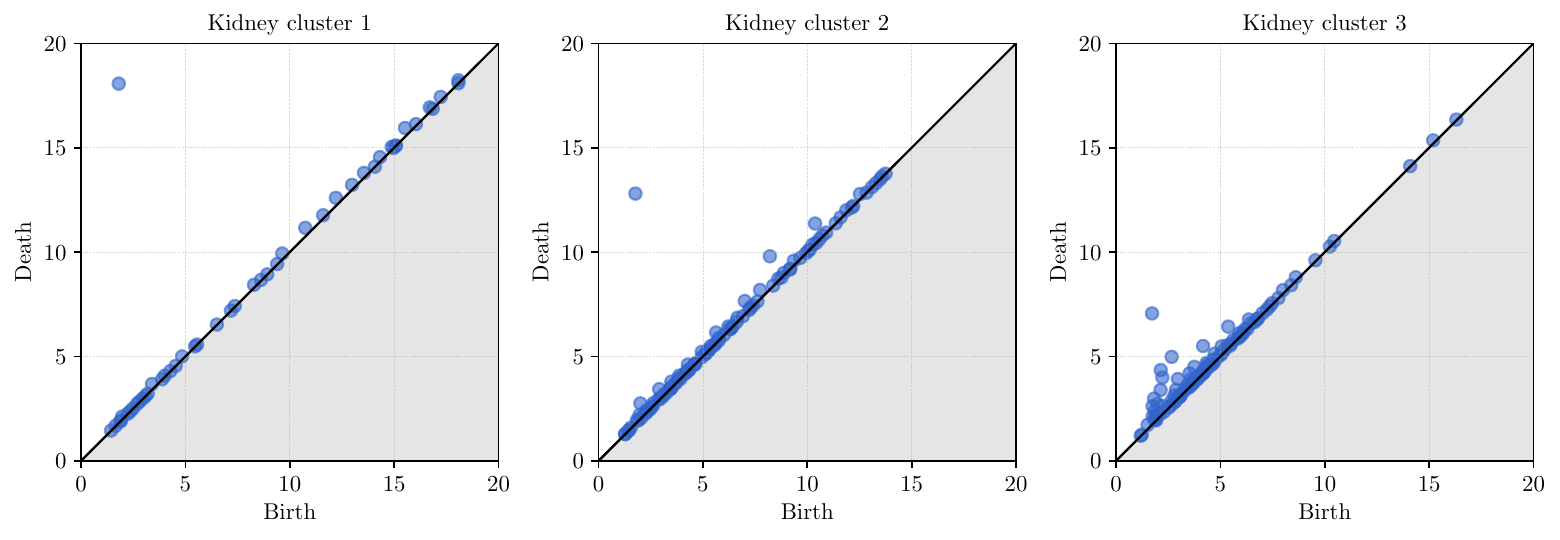}
    \caption{Barycenters of the $H_2$ persistence diagrams within each cluster for brains (top) and kidneys (bottom). Clusters were obtained using \(k\)-medoids clustering of the fused distance \(d^{\mathrm{fus}}\), derived from \(d^{\mathrm{B}}\) and \(d^{\phi}\).}
    \label{fig:bary_br}
\end{figure}

Figure~\ref{fig:tda_br_kd_mds} shows the MDS embeddings of the fused distance with the $k$-medoids cluster. Figure~\ref{fig:bary_br} shows the barycenters of the persistence diagrams within each cluster. Both are discussed in Section~\manualref{4.5} of the main text.

\section{Simulation details for Section~\manualref{4.4}}
\label{sec:SimuDetails}

Each trajectory begins (at $t=1$) as an ellipsoid with semi-axes $a=8$, $b=5.5$, $c=4.0$ along the $x$-, $y$-, and $z$-axes, with points sampled uniformly from its surface. At each step $t=2,\ldots,300$, two spherical cavities are added. The centers of the cavities were sampled from the interior of the ellipsoid. Along the \(x\)-axis, the sampling probability changed linearly from one end of the ellipsoid to the other, with the probability five times higher at one end than the other. Along the \(y\)- and \(z\)-axes, the locations were sampled uniformly. 

\begin{figure}[h!]
    \centering   
    \includegraphics[width=1\linewidth]{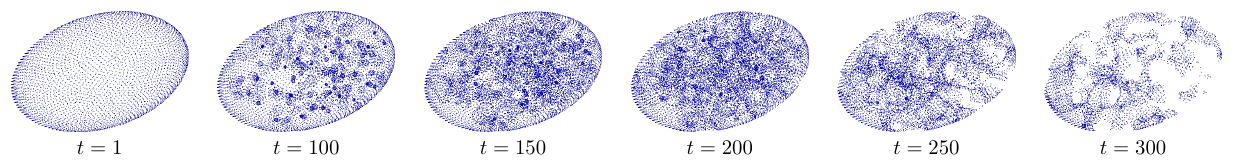}
    \caption{One simulated trajectory shown at selected time points from the full sequence of 300 time points; interior points mark cavity walls.}
    \label{fig:simu_t}
\end{figure}

Once formed, a cavity's center remains fixed, while its radius grows linearly with age, defined as $r(\Delta t) = r_0 + 0.005\,\Delta t$, where $r_0 = 0.1$ and the radius is capped at $1.35$. For each cavity, points are sampled uniformly from its cavity boundary. When two or more cavities overlap, boundary points inside another cavity are removed, so that only the boundary of the merged cavity are retained. When a cavity meets the outer surface of the ellipsoid, surface points inside the cavity and cavity points are removed; cavity points outside the ellipsoid are also removed. The point cloud at time $t$ therefore samples the boundary of the ellipsoid minus the union of its cavities. Figure~\ref{fig:simu_t} shows an example simulated trajectory at selected time points from the full sequence of 300 time points.

\section{Selection of number of clusters}
\label{sec:NumberofClusters}
The number of clusters $k$ was determined using the three criteria described in the main text. The values of these criteria are reported in Table~\ref{tab:pnc} for the registration-based PNC method and in Table~\ref{tab:tda} for the registration-free TDA method.

\begin{table}[ht!]
\centering
\caption{Internal validation results for \(k\)-medoids clustering of the brain data using the PNC geodesic distance matrix.}
\label{tab:pnc}
\begin{tabular}{rrrr}
\toprule
$k$ & Connectivity & Dunn & Silhouette \\
\midrule
2 & 7.175  & 0.064 & 0.643 \\
3 & 4.127  & 0.272 & 0.736 \\
4 & 10.722 & 0.146 & 0.684 \\
\bottomrule
\end{tabular}
\end{table}

\begin{table}[ht!]
\centering
\caption{Internal validation results for \(k\)-medoids clustering using \(d^{\mathrm{fus}}\), derived from \(d^{\mathrm{B}}\) and \(d^{\phi}\).}
\label{tab:tda}
{\setlength{\aboverulesep}{0pt}
\setlength{\belowrulesep}{0pt}
\begin{tabular}{rrrr|rrrr}
\toprule
\multicolumn{4}{c}{Brain} & \multicolumn{4}{|c}{Kidney} \\
\cmidrule(lr){1-4} \cmidrule(lr){5-8}
$k$ & Connectivity & Dunn & Silhouette &
$k$ & Connectivity & Dunn & Silhouette \\
\midrule
2 & 0.211  & 0.391 & 0.674 & 2 & 8.878  & 0.093 & 0.513 \\
3 & 5.451  & 0.250 & 0.500 & 3 & 12.876 & 0.103 & 0.475 \\
4 & 12.919 & 0.099 & 0.467 & 4 & 21.183 & 0.156 & 0.489 \\
\bottomrule
\end{tabular}}
\end{table}

\section{Sensitivity to rank aggregation methods}
\label{sec:SensitivityAgg}

\begin{figure}[h!]
    \centering   
    \includegraphics[width=.85\linewidth]{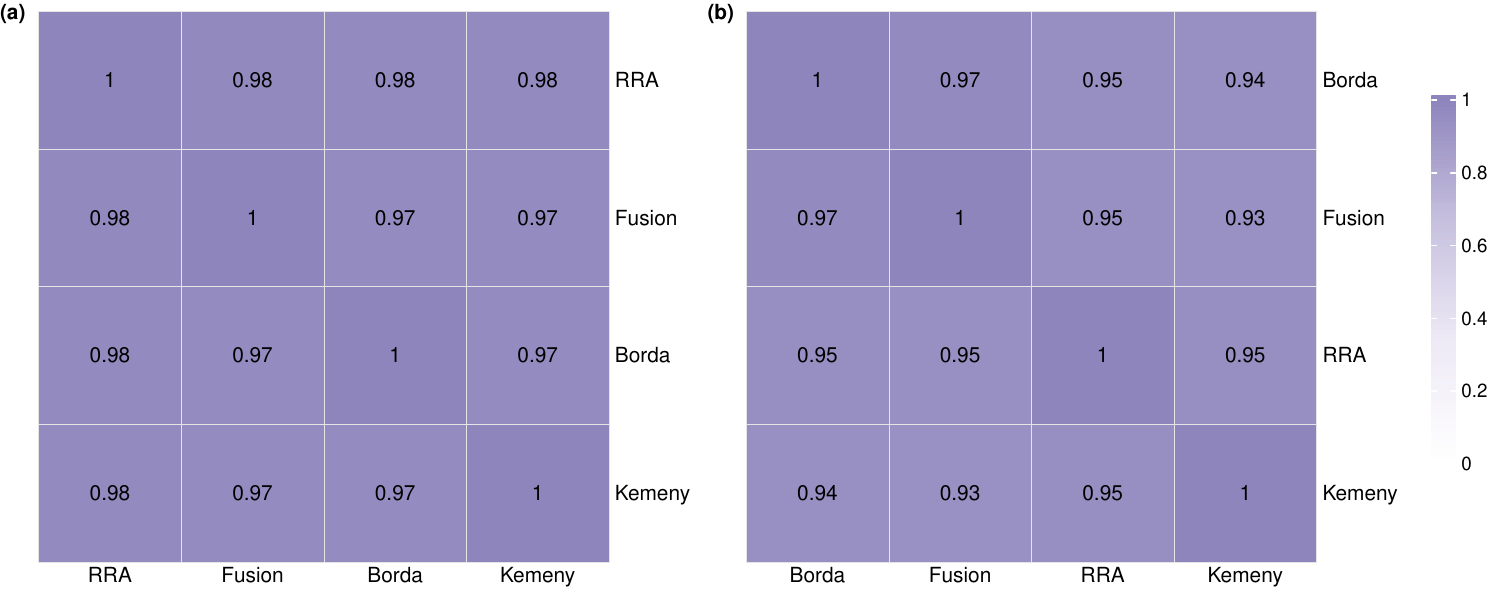}
    \caption{Kendall’s $\tau$ between rankings produced by different rank-aggregation methods for (a) brain samples and (b) kidney samples. Kemeny refers to Kemeny rank aggregation \citep{ConsRank}; Fusion refers to seriation based on fused distance matrices; Borda refers to Borda rank aggregation; and RRA refers to Robust Rank Aggregation \citep{kolde2012robust,RobustRankAggreg}.}
    \label{fig:rank_agg_comp}
\end{figure}

Figure~\ref{fig:rank_agg_comp} reports pairwise Kendall's $\tau$ between the aggregated rankings using Borda, Kemeny, and RRA. All values are at least 0.93, indicating that the aggregated orderings do not depend strongly on the aggregation methods.

\section{Comparison of decomposition severity between left and right kidneys}
\label{sec:LRKidney}

\begin{figure}[h!]
    \centering   
    \includegraphics[width=.7\linewidth]{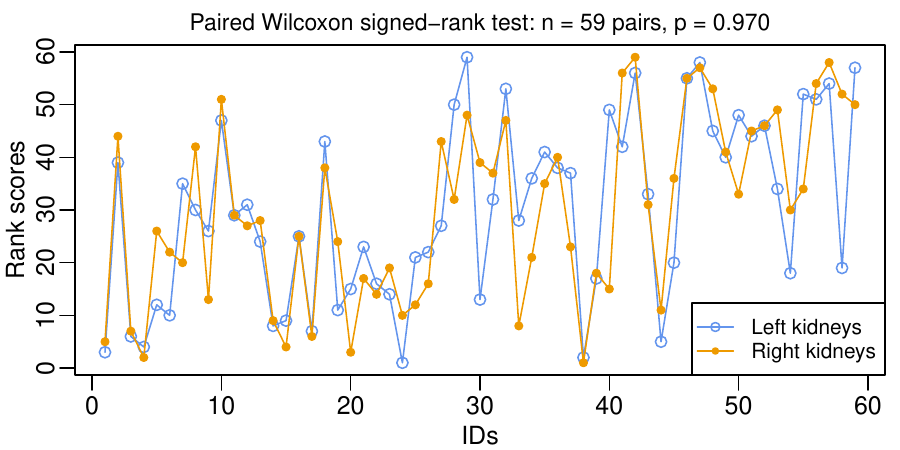}
    \caption{Final rank scores of left and right kidneys from $d^\mathrm{fus}$ and results from paired Wilcoxon signed-rank test.}
    \label{fig:wilcox}
\end{figure}

Rank scores for paired left and kidneys are shown in Fig.~\ref{fig:wilcox}. A two-sided paired Wilcoxon signed-rank test was used to assess whether decomposition severity differed systematically between left and right kidneys from the same individual. No significant difference was detected ($p = 0.970$), indicating no evidence that either side is consistently more or less decomposed than the other.

\end{document}